\documentclass[prd,onecolumn,amsmath,showpacs,nofootinbib,11pt]{revtex4}
\usepackage{graphicx}
\usepackage{float}
\usepackage{dcolumn}
\usepackage{bm}
\begin{document}
\newcommand{\hs}{\hspace*{0.5cm}}
\newcommand{\vs}{\vspace*{0.5cm}}
\newcommand{\be}{\begin{equation}}
\newcommand{\ee}{\end{equation}}
\newcommand{\bea}{\begin{eqnarray}}
\newcommand{\eea}{\end{eqnarray}}
\newcommand{\ben}{\begin{enumerate}}
\newcommand{\een}{\end{enumerate}}
\newcommand{\bde}{\begin{widetext}}
\newcommand{\ede}{\end{widetext}}
\newcommand{\nn}{\nonumber}
\newcommand{\crn}{\nonumber \\}
\newcommand{\Tr}{\mathrm{Tr}}
\newcommand{\non}{\nonumber}
\newcommand{\noi}{\noindent}
\newcommand{\al}{\alpha}
\newcommand{\la}{\lambda}
\newcommand{\bet}{\beta}
\newcommand{\ga}{\gamma}
\newcommand{\va}{\varphi}
\newcommand{\om}{\omega}
\newcommand{\pa}{\partial}
\newcommand{\+}{\dagger}
\newcommand{\fr}{\frac}
\newcommand{\bc}{\begin{center}}
\newcommand{\ec}{\end{center}}
\newcommand{\Ga}{\Gamma}
\newcommand{\de}{\delta}
\newcommand{\De}{\Delta}
\newcommand{\ep}{\epsilon}
\newcommand{\varep}{\varepsilon}
\newcommand{\ka}{\kappa}
\newcommand{\La}{\Lambda}
\newcommand{\si}{\sigma}
\newcommand{\Si}{\Sigma}
\newcommand{\ta}{\tau}
\newcommand{\up}{\upsilon}
\newcommand{\Up}{\Upsilon}
\newcommand{\ze}{\zeta}
\newcommand{\ps}{\psi}
\newcommand{\Ps}{\Psi}
\newcommand{\ph}{\phi}
\newcommand{\vph}{\varphi}
\newcommand{\Ph}{\Phi}
\newcommand{\Om}{\Omega}

\title{Investigation of Dark Matter in the 3-2-3-1 Model}

\author{D. T. Huong}\email{dthuong@iop.vast.ac.vn}\affiliation{Institute of Physics, Vietnam Academy of Science and Technology, 10 Dao Tan, Ba Dinh, Hanoi, Vietnam} 
\author{P. V. Dong}\email{pvdong@iop.vast.ac.vn}\affiliation{Institute of Physics, Vietnam Academy of Science and Technology, 10 Dao Tan, Ba Dinh, Hanoi, Vietnam}
\author{N. T. Duy}\email{ntdem@iop.vast.ac.vn}\affiliation{Institute of Physics, Vietnam Academy of Science and Technology, 10 Dao Tan, Ba Dinh, Hanoi, Vietnam}  
\author{N. T. Nhuan}\email{ntnhuan@grad.iop.vast.ac.vn}\affiliation{Graduate University of Science and Technology, Vietnam Academy of Science and Technology, 18 Hoang Quoc Viet, Cau Giay, Hanoi, Vietnam}
\author{L. D. Thien}\email{ldthien@grad.iop.vast.ac.vn}\affiliation{Graduate University of Science and Technology, Vietnam Academy of Science and Technology, 18 Hoang Quoc Viet, Cau Giay, Hanoi, Vietnam}

\begin{abstract}

We prove that the $SU(3)_C\otimes
SU(2)_L \otimes SU(3)_R\otimes U(1)_X$ (3-2-3-1) gauge model always contains a matter parity $W_P=(-1)^{3(B-L)+2s}$ as conserved residual gauge symmetry, where $B-L=2(\beta T_{8R}+X)$ is a $SU(3)_R\otimes U(1)_X$ charge. Due to the non-Abelian nature of $B-L$, the $W$-odd and $W$-even fields are actually unified in gauge multiplets. We investigate two viable versions for dark matter according to $\beta=\pm1/\sqrt{3}$, where the dark matter candidates can be fermion, scalar, or vector fields.  We figure out the parameter spaces in the allowed regions of the relic density and direct detection cross-sections. Additionally, we examine the neutrino masses induced by the seesaw mechanism along with associated lepton flavor violation processes. The new gauge boson searches at the LEPII and LHC are discussed.    
\end{abstract}

\date{\today}

\maketitle

\section{\label{intro}Introduction}

The standard model is very successful, but it is not a complete theory as failing to address the existence of nonzero small neutrino masses and neutrino mixing \cite{neuos} as well as the presence of dark matter that occupies roundly 26\% mass-energy density of the universe \cite{darkrv}. From a theoretical point of view, the standard model cannot explain the existence of three families and the origin of electroweak parity violation \cite{data}.

Among the standard model's extensions, the minimal left-right symmetric model 
is intriguing, which provides natural explanations for the electroweak parity asymmetry and small neutrino masses~\cite{LR1,NLR}\footnote{See for other seesaw interpretation \cite{ad0}.}. Further, the new physics plays important roles in interpreting the neutral meson mixings and rare meson decays \cite{PheLR} as well as appropriate answer to the $V_{ub}$ problem \cite{vubpro}. However, the model does not give a natural solution for dark matter and family number. 

Indeed, the lightest right-handed neutrino may be tuned to have a mass in keV regime responsible for long-lived warm dark matter. But, they would overpopulate the universe due to relevant gauge interactions which acquires non-standard dilution mechanisms as well as testable strict phenomena \cite{ad1}. Less fine-tuning is cold dark matter scenarios that necessarily add a new field to the model and impose a stabilizing symmetry, typically a matter parity as residual $B-L$ gauge symmetry or just the gauge symmetry like the minimal dark matter model or global symmetries~\cite{ad2}. However, the dark sector of such attempts still remains to be arbitrary, ad hoc included, because it is decoupled from (i.e., commuted with) the normal sector under the gauge symmetry. 

Therefore, we would like to search for a stabilizing mechanism by virtue of a noncommutative $B-L$ gauge symmetry that uniquely determines dark matter from the known normal matter as forming an irreducible gauge multiplet (dark matter, normal matter) by symmetry principles. This is in sharp contrast to the usual global and Abelian $B$, $L$, $B-L$ extensions of the standard model, including the minimal left-right symmetric model. Generally, the dark matter has an anomalous (wrong) $B-L$ number, while the normal matter has a normal $B-L$ number. Furthermore, the breaking of this noncommuting $B-L$ symmetry defines both the seesaw scale that keeps small neutrino masses and the matter parity that stabilizes dark matter. By this proposal, the smallness of neutrino masses and the stability of dark matter are originally connected, in the same nature. However, in the model the dark matter mass is set by another new physics scale that separates the multiplet of dark matter and normal matter, opposite to the usual interpretation \footnote{Comparing to supersymmetry (a spacetime symmetry, by contrast), dark matter (sparticle) and normal matter (particle) are unified in supermultiplet, differ in spin, and are split by supersymmetry breaking. But, the dark matter stabilization is due to $R$-parity---a residual $R$-symmetry having an undefined nature.}.                           

The idea of realizing a noncommuting $B-L$ gauge symmetry (i.e., combining it with electroweak charges in a non-Abelian group) often composes the solution of family number (see, for instance,~\cite{3311dm}). Therefore, there have recently attempted to solve the mentioned last two questions (dark matter and family number) by enlarging the left and/or right weak-isospin groups, i.e. $SU(2)_{L,R}$ \cite{LR331,3L3R, 2L3R,3L3RVE,lrema}. The simplest versions achieved include gauge symmetries, $SU(3)_C\otimes SU(2)_L  \otimes SU(3)_R \otimes U(1)_{X}$ (3-2-3-1) and $SU(3)_C\otimes SU(3)_L  \otimes SU(2)_R \otimes U(1)_{X}$ (3-3-2-1), respectively. However, the former is somewhat simpler than the latter due to its left-handed sector remaining as the standard model, to be investigated further in this work. A predictive feature of such left-right gauge extension is that the $B-L$ charge of new fields is determined by their electric charge, $Q=T_{3L}+T_{3R}+\fr{1}{2}(B-L)$. For instance, new $SU(2)_{L,R}$ singlets have $B-L=2Q$, which differs from that of ordinary particles. In fact, the most new fields have a wrong $B-L$ charge, if they have a usual electric charge. The highlight of the present model is that all the existing issues can be manifestly described by gauge principles---a gauge completion.     

Indeed, the 3-2-3-1 model explains the family number by $SU(3)_R$ anomaly cancellation, analogous to the 3-3-1 model \cite{331}. It already provides the neutrino masses via a seesaw mechanism similar to the minimal left-right symmetric model \cite{LR1,NLR}. Unlike the conventional dark matter theories, the 3-2-3-1 model encloses and treats $B-L=2(\beta T_{8R}+X)$ as a noncommuting gauge charge of $SU(3)_R\otimes U(1)_X$. Due to this fact, the model naturally accommodates dark matter as component fields that complete the $SU(3)_R$ multiplets, e.g. $3=2\oplus 1 = (N,N,D)$ or $(D,D,N)$, $6=3\oplus 2\oplus 1 = (N,N,N,D,D,N)$, and $8=3\oplus 2\oplus 2^*\oplus 1=(N,N,N,D,D,D^*,D^*,N)$, under $SU(2)_R$, where $N$ and $D$ refer to a normal field and dark field, respectively. The $B-L$-charged scalar field breaks the 3-2-3-1 symmetry, defining both the seesaw scale as the scalar vacuum value producing small neutrino masses and the matter parity $W_P=(-1)^{3(B-L)+2s}$ as residual $SU(3)_R\otimes U(1)_X$ gauge symmetry responsible for dark matter stability. The $D$ fields (having wrong $B-L$ number) are $W$-odd, whereas the $N$ fields (having normal $B-L$ number) are $W$-even~\footnote{Recall that these opposite parities are actually arranged in gauge multiplets reflecting non-Abelian $B-L$ symmetry.}. The lightest $W$-odd particle (LWP) is a dark matter candidate \footnote{if it is electrically- and color-neutral and has a correct density.}, stabilized by the matter parity conservation. Additionally, in the model the tree-level FCNCs arise due to the discrimination of right-handed quark families under the gauge symmetry \cite{2L3R}. This easily addresses the issues of the neutral meson mixings and rare meson decays~\cite{bsmumu}. The new physics is mostly predicted in the TeV region, which may be searched at the current colliders.
 
The rest of this paper is organized as follows: In Sec. \ref{model}, we give a review of the model, examining the matter parity, dark matter versions and candidates. Additionally, we discuss the existing bounds and necessary conditions for the scalar potential. Sec. \ref{nmflv} presents the neutrino mass generation and lepton flavor violation processes. Sec. \ref{collisionb} studies the new gauge bosons at the colliders. Sec. \ref{VCT} calculates dark matter observables. We make conclusions in Sec. \ref{Con}.

\section{\label{model}A review of the model}

This section summarizes work already done in \cite{2L3R}. But, the gauge symmetry breaking and the origin of the matter parity and dark matter stability are extensively discussed.  

\subsection{Gauge symmetry and particle content}

As stated, the gauge symmetry is given by 
 \be SU(3)_C \otimes SU(2)_L \otimes SU(3)_R \otimes
 U(1)_X,\ee where the hypercharge is enlarged as $U(1)_Y\rightarrow SU(3)_R\otimes U(1)_X$, which will result in the family number, neutrino mass, and dark matter due to the nature of the extended group. 
 
Since the gauge symmetry contains those of the minimal left-right symmetry and the standard model, the electric charge, hypercharge and baryon-minus-lepton charge are embedded as
\be Q = T_{3L}+T_{3R}+\beta T_{8R}+X, \hs Y=T_{3R}+\beta T_{8R}+X,\hs \fr 1 2 (B-L)  =  \beta T_{8R}+X, \label{electric1} \ee
where $T_{aL}$ ($a=1,2,3$), $T_{iR}$
($i=1,2,3,...,8$),  and $X$ are $SU(2)_L$, $SU(3)_R$, and $U(1)_X$ charges, respectively. $Q$, $Y$, and $B-L$ act as non-Abelian charges, not commuting with the gauge symmetry. The coefficient $\beta$ is arbitrary. Particularly, $Q$ and $B-L$ satisfy           
\bea
&& \left[Q, T_4\pm i T_5\right] = \mp q (T_4\pm i T_5),\hs  \left[Q, T_6\pm i T_7\right] = \mp (1+q) (T_6\pm i T_7),\\
&& \left[B-L, T_4\pm i T_5\right] = \mp (1+2q) (T_4\pm i T_5),\hs \left[B-L, T_6\pm i T_7\right] = \mp (1+2q) (T_6\pm i T_7), \eea
where $q\equiv -(1+\sqrt{3}\beta)/2$ will define the electric charge and $B-L$ for new particles. 

The fermions transform under the gauge symmetry as 
 \bea
\psi_{aL}=\left(%
\begin{array}{c}
  \nu_{aL} \\
  e_{aL}\\
\end{array}%
\right)\sim \left(1,2,1,-\frac{1}{2}\right), \hs \hs \psi_{aR}=\left(%
\begin{array}{c}
  \nu_{aR} \\
 e_{aR} \\
  E_{aR}^q \\
\end{array}%
\right)\sim \left(1,1,3,\frac{q-1}{3}\right),
 \eea
 \bea
 Q_{\al L} =\left(%
\begin{array}{c}
  u_{\al L} \\
  d_{\al L} \\
\end{array}%
\right)\sim \left(3,2,1,\frac{1}{6}\right), \hs \hs Q_{\al R}=\left(%
\begin{array}{c}
  d_{\al R} \\
  -u_{\al R} \\
  J_{\al R}^{-q-\frac{1}{3}} \\
\end{array}%
\right)\sim \left(3,1,3^*,-\frac{q}{3}\right),
 \eea
 \bea
Q_{3L}= \left(%
\begin{array}{c}
  u_{3L} \\
  d_{3L} \\
\end{array}%
\right) \sim \left(3,2,1,\frac{1}{6}\right), \hs \hs Q_{3R}= \left(%
\begin{array}{c}
  u_{3R} \\
  d_{3R} \\
  J^{q+\frac{2}{3}}_{3R} \\
\end{array}%
\right)\sim \left(3,1,3,\frac{q+1}{3}\right),
 \eea
 \bea
 E^q_{aL}\sim (1,1,1,q), \hs J^{-q-\fr 1 3}_{\al L} \sim
 \left(3,1,1,-q-\fr{1}{3}\right), \hs J^{q+\fr 2 3}_{3L}\sim \left(3,1,1,q+\fr{2}{3}\right),
 \eea
where $a=1,2,3$ and $\al=1,2$ are family indices. $\nu_R$, $E$, and $J$ are new particles included to complete the representations. To cancel $[SU(3)_R]^3$ anomaly, the number of $SU(3)_R$ fermion triplets must equal the number of $SU(3)_R$ fermion antitriplets. Therefore, the family number must match that of color, and the third quark family must be arranged differently from the first two. It is verified that all other anomalies are cancelled too. Note that after symmetry breaking the $SU(3)_R$ antitriplets and triplets decompose under $SU(2)_R$ as $3^*=2^*\oplus 1$ and $3=2\oplus 1$, i.e. $(d_{\al R}\ -u_{\al R}\ J_{\al R})^T=(d_{\al R}\ -u_{\al R})^T\oplus J_{\al R}$ and $(u_{3R}\ d_{3R}\ J_{3R})^T=(u_{3R}\ d_{3R})^T\oplus J_{3R}$, respectively. Since every $SU(2)_R$ representation is real, the antidoublets $(d_{\al R}\ -u_{\al R})^T$ are equivalent to the doublets $(u_{\al R}\ d_{\al R})^T$. Strictly speaking, since $2^*=i\sigma_2 2$, we derive $(u_{\al R}\ d_{\al R})^T=-i\sigma_2 (d_{\al R}\ -u_{\al R})^T$, which acquire the same $SU(2)_R$ quantum number as $(u_{3R}\ d_{3R})^T$. Using Eq. (\ref{electric1}), it is easily checked that $d_{\al R}$ $(u_{\al R})$ and $d_{3 R}$ ($u_{3 R}$) have the same $Y$ and $Q$, whereas all the ordinary quarks have the same $B-L=1/3$, where note that $T_{iR}=\la_{i}/2$ for triplets while $T_{iR}=-\la^*_{i}/2$ for antitriplets. All the new and usual quarks have the same $SU(3)_C$ quantum number, 3, which is unbroken.                 

To break the gauge symmetry and generate the particle masses appropriately, the scalar content is introduced as
\bea
S &=& \left(%
\begin{array}{ccc}
	S_{11}^0 & S_{12}^+ & S_{13}^{-q} \\
	S_{21}^- & S_{22}^0 & S_{23}^{-q-1} \\
\end{array}%
\right) \sim \left(1,2,3^*,-\frac{2q+1}{6}\right),\\ 
\phi &=& \left(%
\begin{array}{c}
	\phi_1^{-q} \\
	\phi_2^{-q-1} \\
	\phi_3^0 \\
\end{array}%
\right)\sim \left(1,1,3,-\frac{2q+1}{3}\right), \\  
\Xi &=& \left(%
\begin{array}{ccc}
	\Xi^0_{11} &\frac{ \Xi_{12}^-}{\sqrt{2}} &\frac{ \Xi_{13}^q }{\sqrt{2}}\\
	\frac{ \Xi_{12}^-}{\sqrt{2}} & \Xi_{22}^{--} & \frac{\Xi_{23}^{q-1}}{\sqrt{2}} \\
	\frac{\Xi_{13}^q}{\sqrt{2}} & \frac{\Xi_{23}^{q-1}}{\sqrt{2}} & \Xi_{33}^{2q}\\
\end{array}%
\right)\sim \left(1,1,6,\frac{2(q-1)}{3}\right),
\eea
with vacuum expectation values (VEVs), \bea \langle S\rangle  &=&\frac{1}{\sqrt{2}}\left(%
\begin{array}{ccc}
	u & 0 & 0 \\
	0 & v & 0 \\
\end{array}%
\right), \hs  \langle \phi\rangle =\frac{1}{\sqrt{2}}\left(%
\begin{array}{c}
	0 \\
	0 \\
	w\\
\end{array}%
\right), \hs  \langle \Xi\rangle =\frac{1}{\sqrt{2}}\left(%
\begin{array}{ccc}
	\Lambda & 0 & 0 \\
	0 & 0 & 0 \\
	0 & 0 & 0 \\
\end{array}%
\right).\eea  

The fields' superscript is electric charge. The $B-L$ charge is given in Table~{\ref{tb1}}, in which the gauge fields were specified in \cite{2L3R}. We can divide particles into two classes: normal particles include the standard model and new particles carrying normal $B-L$ charge or differing from that by even unit, whereas wrong particles are those having abnormal $B-L$ charge which depends on $q$. The wrong and normal particles are manifestly unified in the right gauge multiplets, for instance, lepton $(\nu\ e\ E)$, quark $(u\ d\ J)$ or $(d\ -u\ J)$, gauge boson $(W^\pm\ Z\ Z'\ X^{\pm q}\ Y^{\pm(1+q)})$, and so on. 
\begin{table}[h]
	\begin{center}
		\begin{tabular}{c|cccccccccccc}
			\hline \hline
			Particle & $\nu_{a}$ &  $e_a$ & $E_{a}^q$ & $u_a$ & $d_a$  & $J_{\al}^{-q-\fr{1}{3}}$ & $J_3^{q+\fr{2}{3}}$ & $\ph_1^{-q}$ & $\phi_2^{-(q+1)}$ & $\phi_3^0$ \\ \hline
			$B-L$ & $-1$ & $-1$ & $2q$ & $\frac{1}{3}$ & $\frac{1}{3} $& $-\frac{2(1+3q)}{3}$ & $\frac{2(2+3q)}{3}$&$-(1+2q)$ & $-(1+2q)$ & $0$  \\ \hline
			$W_P$ & 1 & $1$ & $P^+$ & 1 & $1$ & $P^-$ & $P^+$ & $P^-$ & $P^-$ & 1 \\
			\hline \hline
			Particle &$S_{11}^0$ & $S_{12}^+$  & $S_{13}^{-q}$ & $S_{21}^-$ & $S_{22}^0$ & $S_{23}^{-1-q}$&$\Xi_{11}^0$&$\Xi_{12}^-$&$\Xi_{13}^q$& $\Xi_{22}^{--}$  \\ \hline
			$B-L$&  $0$   & $0$ & $-(1+2q)$ & $0$ & $0$ & $-(1+2q)$&$-2$ &$-2$& $2q-1$ &$-2$  \\  \hline
			$W_P$ & 1 & $1$ & $P^-$ & 1 & 1 & $P^-$ & $1$ & $1$ & $P^+$ & 1 \\
			\hline\hline
			Particle &  $\Xi_{23}^{q-1}$& $\Xi_{33}^{2q}$&$A$&$Z_{L,R }$&$Z_{R}^\prime$&$W^\pm_{L,R}$&$X_{R}^q$& $X_{R}^{-q}$&$Y_{R}^{q+1}$&$Y_{R}^{-(q+1)}$\\ \hline
			$B-L$   & $2q-1$ & $4q$ &$0$& $0$ & $0$&$0$&$1+2q$& $-(1+2q)$&$1+2q$&$-(1+2q)$\\ \hline
			$W_P$ & $P^+$ & $P^+P^+$  & 1 & 1 & 1 & 1 & $P^+$ & $P^-$ & $P^+$ & $P^-$\\
			\hline\hline 
		\end{tabular}
	\end{center}
	\caption{\label{tb1} $B-L$ charge and $W$-parity for the model's particles.}
\end{table}

\subsection{Symmetry breaking and $W$-parity}

The spontaneous symmetry breaking is implemented through three possible ways. The first way assumes $w \gg \La \gg u,v$, and the gauge symmetry is broken as
\bc
\begin{tabular}{c}  
	$SU(3)_C\otimes SU(2)_L\otimes SU(3)_R \otimes U(1)_X$\\
	$  \downarrow w $\\
	$SU(3)_C \otimes SU(2)_L \otimes SU(2)_R \otimes U(1)_{B-L}$\\
	$\downarrow \La $\\ 
	$SU(3)_C\otimes SU(2)_L \otimes U(1)_Y\otimes W_P$ \\ $\downarrow u,v $\\ $SU(3)_C \otimes U(1)_Q \otimes W_P$. 
\end{tabular} \label{pv2}
\ec

Note that $SU(2)_R$ includes $T_{1R}, T_{2R},T_{3R}$. $B-L$ and $Y$ as given commute with $SU(2)_{L,R}$ and $SU(2)_{L}$, respectively. $W_P\equiv e^{i\omega(B-L)}$ is the residual symmetry of $B-L$ which conserves the vacuum, $W_P\La=\La$. We deduce $\omega=k\pi$ for $k$ integer, and thus $W_P=(-1)^{k(B-L)}$. Among the survival transformations, considering $k=3$ and conveniently multiplying the spin-parity $(-1)^{2s}$ as conserved by the Lorentz symmetry, we obtain the matter parity 
\be W_P=(-1)^{3(B-L)+2s}=(-1)^{6(\beta T_{8R}+X)+2s}.\ee 

Another consequence of this symmetry breaking scheme is that the world may start from an explicit left-right asymmetric phase, translating to an intermediate left-right symmetric phase, and going down the electroweak phase by spontaneous parity breaking.      

The second way assumes  
$\La \gg  w \gg u,v$, and the gauge symmetry is broken as
\bc
\begin{tabular}{c}  
	$SU(3)_C\otimes SU(2)_L\otimes SU(3)_R \otimes U(1)_X$\\
	$  \downarrow \Lambda$\\
	$SU(3)_C\otimes SU(2)_L \otimes SU(2)_{R^\prime}\otimes U(1)_{X^\prime} \otimes W'_P$\\
	$\downarrow w$\\ 
	$SU(3)_C\otimes SU(2)_L\otimes U(1)_Y\otimes W_P$\\ $\downarrow u,v$ \\ $SU(3)_C \otimes U(1)_{Q} \otimes W_P$.
\end{tabular} 
\ec
$SU(2)_{R^\prime}$ contains three generators, $T_{6R}, T_{7R}, \fr{1}{2}\left(\sqrt{3}T_{8R}-T_{3R}\right)$, and $U(1)_{X^\prime}$ is $X^\prime =\fr 1 4 (\sqrt{3}+\beta)(T_{8R}+\sqrt{3}T_{3R})+X$. $W'_P=(-1)^{\fr{3\beta}{2}(\sqrt{3}T_{3R}+T_{8R})+6X}$ is a discrete symmetry orthogonal to $X'$, defined by $\La$ (note that $B-L$ is not factorized at this stage), and $W_P$ takes the normal one (after multiplying the spin parity) determined by $w$. Since $SU(2)_L$ is not interchanged to $SU(2)_{R'}$, there is no left-right symmetric phase for this scheme. In other words, this way breaks the gauge symmetry to the alternative left-right model, rather than the left-right.

The last case is $w \sim \La$, and the gauge symmetry is broken as
\bc
\begin{tabular}{c}  
	$SU(3)_C\otimes SU(2)_L\otimes SU(3)_R \otimes U(1)_X$\\
	$  \downarrow w, \La $\\
	$SU(3)_C \otimes SU(2)_L \otimes U(1)_Y \otimes W_P$
	 \\ $\downarrow u,v $\\ $SU(3)_C \otimes U(1)_Q \otimes W_P$. 
\end{tabular} \label{pv3}
\ec
It is easily shown that $B-L$ commutes with the standard model symmetry and its remnant is $W_P$ defined by $\La$ vacuum. This case does not recognize the left-right symmetric phase.   

Therefore, every symmetry breaking scheme leads to the matter parity $W_P$ as a residual gauge symmetry, which is not commuted with the beginning gauge symmetry. Its value is listed in Table \ref{tb1}. The normal particles have $W_P=1$. The wrong particles have $W_P=P^+$ or $P^-$, where $P^\pm = (-1)^{\pm(6q+1)}\neq 1$ is non-trivial if $q\neq \fr{2m-1}{6}=\pm \fr{1}{6}, \pm \fr{1}{2}, \pm \fr{5}{6}, \pm\fr{7}{6},...$ for 
all $m$ integer. This assumption is natural, since those values of the electric charge are unlikely. For instance, if $q$ takes normal charges, i.e. $q=m/3$, then $P^\pm=-1$ and the wrong particles become odd fields.  

The $W_P$ conservation implies that wrong particles always couple in pairs or self-couple. Indeed, consider an interaction consisting of $x$ $P^+$ fields and $y$ $P^-$ fields for $x,y$ integers. $W_P$ is conserved, leading to $(P^+)^x(P^-)^y=(-1)^{(6q+1)(x-y)}=1$, thus $x=y$ for arbitrary $q$. $P^+$ and $P^-$ always appear in pairs. This also applies for $(P^+)^2$ and $(P^-)^2$ fields. If an interaction contains either $(P^+)^2$ or $(P^-)^2$ field, it has two other either $P^-$ or $P^+$ fields, respectively, leading to the self-couple of three $W$-fields. Therefore, the lightest wrong particle (LWP) is stabilized responsible for dark matter if it carries no electric and color charges.  We have three versions for dark matter corresponding to $q=0,-1,+1$ or $\beta=-1/\sqrt{3},1/\sqrt{3},-\sqrt{3}$, respectively. However, the version $q=1$ is ruled out by matching the gauge couplings as shown below. 

The version $q=0$ includes dark matter candidates as a fermion combined of $E_{1,2,3}$, a scalar combined of $\ph_1, S_{13}, \Xi_{13}$, or a gauge boson $X_{R}$. The version $q=-1$ has dark matter candidates as a gauge boson $Y_{R}$ or a scalar combined of $\ph_2, S_{23}$. As studied in \cite{2L3R}, one combination of $\phi_2,S_{23}$ is the Goldstone boson of $Y_R$, the corresponding candidate is only $H_8=(v\phi_2+wS_{23})/\sqrt{v^2+w^2}$. Similarly, one combination of $\phi_1,S_{13},\Xi_{13}$ is the Goldstone boson of $X_R$, the relevant candidates are $H_6\simeq (u\phi_1+wS_{13})/\sqrt{u^2+w^2}$ and $H_7\simeq (w\phi_1-uS_{13})/\sqrt{u^2+w^2}$. The masses of $E,X_R,Y_R,H_{6,7,8}$ are proportional to $w$ and/or $\La$, which should be radically larger than the weak scale.       

\subsection{Existing constraints}

Let vectors $A_{aL\mu}$, $A_{iR\mu}$, and $B_\mu$ couple to $T_{aL}$, $T_{iR}$, and $X$ in the covariant derivative according to the coupling constants $g_{L,R,X}$ respectively, and denote $t_R\equiv g_R/g_L$, $t_X\equiv g_X/g_L$ \cite{2L3R}. The new gauge bosons  
$X_{R}^{\pm q} = (A_{4R}\pm i A_{5 R})/\sqrt{2},\ Y_{R}^{\pm (q+1)} = (A_{6R}\pm i A_{7 R})/\sqrt{2}$ possess masses $m_{X_R}\simeq \fr{g_R}{2}\sqrt{w^2+2\La^2}$, $m_{Y_R}\simeq \fr{g_R}{2}w$ and decoupled, whereas $W_{L}^\pm = (A_{1L }\mp iA_{2L})/\sqrt{2},\ W_{R}^\pm = (A_{1R }\mp i A_{2R })/\sqrt{2}$ mix, which yield eigenstates $W_{1} =c_\xi W_{L} -s_\xi W_{R},\ W_{2} =s_\xi W_{L} +c_\xi W_{R}$, with the mixing angle $|\xi|\ll 1$ and $m_{W_1} \simeq \fr{g_L}{2}\sqrt{u^2+v^2}$, $m_{W_2}\simeq \fr{g_R}{\sqrt{2}}\La$, where $W_1$ is analogous to the standard model while $W_2$ is new.   
The photon field $A =s_WA_{3L}+c_W\left(\fr{t_W}{t_R}A_{3R}+\beta \fr{t_W}{t_R}A_{8R}+\fr{t_W}{t_X}B\right)$ is massless eigenstate, while the standard model $Z$ boson $Z = c_WA_{3L}-s_W\left(\fr{t_W}{t_R}A_{3R}+\beta \fr{t_W}{t_R}A_{8R}+\fr{t_W}{t_X}B\right)$ slightly mixes with the heavy states $Z_R, Z'_R$, given orthogonally to the field parenthesized in $A,Z$, via the mixing parameters $|\epsilon_{1,2}|\ll 1$.\footnote{Here, $Z_R=[-(t^2_R+\beta^2 t^2_X)A_{3R}+t_X(\beta t_X A_{8R}+t_R B)]/\sqrt{(t^2_R+\beta^2 t^2_X)[t^2_R+(1+\beta^2)t^2_X]}$ and $Z_R^\prime=(t_R A_{8R}-\beta t_X B)/\sqrt{t^2_R+\beta^2 t^2_X}$ finitely mix, which yield physical states $\mathcal{Z}_1=c_\epsilon Z'_R-s_\epsilon Z_R$ and $\mathcal{Z}'_1=s_\epsilon Z'_R+c_\epsilon Z_R$, with the mixing angle $\epsilon$ and their masses dependent on $w,\La$ \cite{2L3R}.}

The VEVs $w,\La$ break the new symmetries and give the masses for new particles, while $u,v$ break the standard model symmetry and provide the masses for ordinary particles. For consistency, we impose $u,v \ll w,\La$. Additionally, the $W$ mass implies $u^2+v^2 \simeq(246\ \mathrm{GeV})^2$.
  
Due to the mixings of $W,Z$ with the respective new gauge bosons, the $\rho$-parameter $\rho=m^2_{W}/c^2_Wm^2_{Z}$ as well as the well-measured couplings of $W,Z$ with fermions are modified through $\xi,\epsilon_{1,2}$ \cite{2L3R}. Fitting to the data, the new physics scales, assuming $w=\La$, take lower bounds in several TeV, for instance, $\La>2.1$ TeV for $\beta=1/\sqrt{3}$ (or $q=-1$) and $\La>3.9$ TeV for $\beta=-1/\sqrt{3}$ (or $q=0$). Note that only the upper bound for $\Delta\rho$ is taken into account, which differs from \cite{2L3R}.     

As presented in \cite{2L3R}, matching the gauge couplings leads to 
\be s^2_W=\fr{t^2_R t^2_X}{t^2_R+t^2_X(1+\beta^2+t^2_R)}<\fr{t^2_R}{1+\beta^2+t^2_R},\ee where recall $t_R=g_R/g_L$, $t_X=g_X/g_L$, and that $g_{L,R,X}$ are $SU(2)_L$, $SU(3)_R$, and $U(1)_X$ couplings, respectively. Taking $t_R=1$ as motivated/protected by the left-right symmetry, we have $s^2_W<1/(2+\beta^2)$, thus $-1.822< q< 0.822$, where note that $\beta=-(1+2q)/\sqrt{3}$ and $s^2_W=0.231$. Comparing to the previous section, there are only two dark matter versions for $q=0,-1$. 

When the energy scale increases, $g_{L,R}$ slightly change, while $g_X$ significantly rises. A Landau pole $M$ at which $s^2_W(M)=1/(2+\beta^2)$ or $g_X(M)=\infty$ may result, depending on $q$, where we set $t_R(M)=1$ for simplicity. Of course, the model works only if $\La,w<M$. For instance, the Landau pole approaches the weak scale if $q$ tends to either of its bounds (these cases should be ruled out by other contraints), and the Landau pole is $M\sim 10$ TeV for $q=-1/2$ or $\beta=0$. Further, the Landau poles for the dark matter versions $q=0,-1$ are actually larger than the Planck scale. 

We would like to emphasize that the source of FCNCs is due to the third right-handed quark multiplet ($Q_{3R}$) transforming differently from the first two ($Q_{\al R}$), i.e. a result of the non-universal fermion families \cite{2L3R}. Thus, the tree-level FCNCs occur via both gauge and Yukawa interactions, with the relevant couplings derived as \cite{2L3R}
\bea
\mathcal{L}_{\mathrm{FCNC}}=\bar{d}_{iL}^\prime \Gamma_{ij}^d d^\prime_{jR}H_2+\bar{u}^\prime_{iL}\Gamma^u_{ij}u^\prime_{jR}H_2 +H.c. -\Theta_{ij}^{Z_R^\prime}\bar{q}_{iR}^\prime \ga^\mu q^\prime_{jR} Z^\prime_{R \mu}, \label{FCNC}
\eea
where $\Gamma^u, \Gamma^d, \Theta^{Z_R^\prime}$ are the couplings that depend only on the ordinary quark mixing matrix elements of both the left and right sectors and the VEVs $(u,v)$. There is no mixing between the ordinary and new quarks due to the matter-parity conservation. It means the interactions of the FCNCs with $H_2,Z'_R$ do not depend on the way of the symmetry breaking, but the amplitudes of the induced effective FCNC interactions do, set by $H_2,Z'_R$ masses. Eq. (\ref{FCNC}) modifies the neutral meson mass differences, $\Delta m_K, \Delta m_{B_s}, \Delta m_{B_d}$, and thus constrains $w$ and $\La$ in a few TeV, in agreement with those from the $\rho$ and mixing parameters \cite{2L3R}. 

All the analyses have been presented with the assumption $w\sim\La$, which is appropriate to the third way of the symmetry breaking. When either $w\gg \La$ (the first way symmetry breaking) or $\La\gg w$ (the second way symmetry breaking, preferred in the current work), all the above bounds simply apply for the corresponding lower scale, with slightly changed numerical-values. In these cases, the higher scale gives no contribution.       

The scalar potential that is invariant under the gauge symmetry and renormalizable is    
\bea V_{\mathrm{scalar}} &= &\mu_S^2 \Tr(S^\dag S)+\la_{1S}[\Tr (S^\dag
S)]^2+\la_{2S}\Tr(S^\dag SS^\dag S) +\mu_\Xi^2\Tr(\Xi^\dag \Xi)\crn 
&& +\la_{1\Xi}[ \Tr(\Xi^\dag \Xi)]^2 +\la_{2 \Xi}\Tr( \Xi^\dag \Xi \Xi^\dag \Xi)+\mu_\phi^2 \phi^\dag \phi +\la_\phi (\phi^\dag \phi)^2\crn
&& +\la_1(\phi^\dag S^\dag S\phi)+\la_2\Tr(S^\dag S \Xi \Xi^\dag)+\la_3(\phi^\dag\Xi\Xi^\dag\phi)+\la_4
(\phi^\dag \phi) \Tr(S^\dag S)\crn
&& +\la_5 (\phi^\dag \phi) \Tr(\Xi^\dag \Xi)+ \la_6\Tr(\Xi^\dag \Xi) \Tr(S^\dag
S) +  (f S\phi^*S+ H.c.),  \label{232} \eea
where the potential parameters are defined similarly to \cite{2L3R}. The necessary conditions for the scalar potential to be bounded from below as well as to induce the gauge symmetry breaking are  
\bea
\la_{1S}+\la_{2S}>0, \hs  \la_{1\Xi}+\la_{2\Xi} >0,  \hs \la_{\phi}>0,\hs \mu_S^2<0,\hs  \mu_{\Xi}^2<0,\hs  \mu_{\phi}^2<0.\eea Additionally, we have four relations from the potential minimization, which imply $f\sim w,\La$, and that all the Higgs masses have to be positive \cite{2L3R}. Expand the neutral scalar fields around their VEVs, $S_{11}=(u+S_1+iA_1)/\sqrt{2}$, $S_{22}=(v+S_2+iA_2)/\sqrt{2}$, $\phi_3=(w+S_3+iA_3)/\sqrt{2}$, and $\Xi_{11}=(\La+S_4+iA_4)/\sqrt{2}$. The states $S_{1,2,3,4}$ mix, but using the approximation, $(u,v)^2/(w,\La,f)^2\ll 1$, the model contains only a light (CP-even) neutral scalar field, $H_1\simeq (uS_1+vS_2)/\sqrt{u^2+v^2}$, to be identified as the standard model Higgs boson.\footnote{Besides, the model includes eleven new heavy Higgs bosons, the neutral $H_{2}= (-vS_1+uS_2)/\sqrt{u^2+v^2}$, $H_{3}=c_\varphi S_3-s_{\varphi} S_4$, $H_4=s_{\varphi} S_3+c_{\varphi} S_4$, $\mathcal{A}=[w(vA_1+uA_2)-uvA_3]/\sqrt{w^2(u^2+v^2)+u^2v^2}$, and the charged $H^{\pm}_5,\ H^{\pm q}_{6,7},\ H^{\pm(q+1)}_8,\ \Xi^{\pm\pm}_{22},\ \Xi^{\pm(q-1)}_{23},\ \Xi^{\pm 2q}_{33}$, as well as eleven Goldstone boson modes, where the mixing angle, the physical states, and their masses can be explicitly found in \cite{2L3R}.} The relevant Higgs mass is constrained by~\cite{data}     
\bea m_{H_1} &\simeq & \sqrt{2(\lambda_{1S}+\lambda_{2S})u^2-\lambda_{2S}v^2}\simeq 125\ \mathrm{GeV}. \label{neutralHiggs}   \eea

\section{\label{nmflv} Neutrino mass and lepton flavor violation}

The Yukawa interactions of leptons are given by \cite{2L3R}
\bea
\mathcal{L}\supset h_{ab}^l \bar{\Psi}_{aL}S \Psi_{bR}+h_{ab}^E \bar{E}_{aL}\phi^\dag \Psi_{bR}+h_{ab}^R \bar{\Psi}^c_{aR}\Xi^\dag \Psi_{bR}+H.c.
\eea
After the symmetry breaking, the charged leptons $(l,E)$ gain appropriate masses,
\be [m_l]_{ab}=-h^l_{ab} \fr{v}{\sqrt{2}},\hs [m_E]_{ab}=-h^E_{ab}\fr{w}{\sqrt{2}},\ee proportional to the weak and large scales, respectively. 

The neutral leptons get Dirac masses via $u$ and right-handed Majorana masses via $\La$, given in the basis $(\nu_L, \nu_R^c)$ as follows
\bea
M_\nu= -\frac{1}{\sqrt{2}}  \left( \begin{array}{cc}
	       0         &     h^l u     \\
	(h^{l})^T u & 2 h^R \Lambda
\end{array}\right).
\eea
Because of $u \ll \Lambda$, the type I seesaw mechanism applies and the active neutrinos ($\sim \nu_L$) obtain small Majorana masses as
\bea
m_{\nu}\simeq \fr{u^2}{2\sqrt{2}\La} h^l (h^{R})^{-1} (h^l)^T.
\eea
By contrast, the sterile neutrinos ($\sim \nu_R$) have large Majorana masses, $m^R_{\nu} \simeq -\sqrt{2} h^R \Lambda$, in the $B-L$ breaking scale. 

Using $h^l=-\sqrt{2}m_l/v$ and $m_\nu\sim 0.1$ eV \cite{data}, we evaluate \be h^R\sim \fr{1}{\sqrt{2}}\left(\fr{u}{v}\right)^2\left(\fr{m_l}{\mathrm{GeV}}\right)^2\fr{10^{10}\ \mathrm{GeV}}{\La}.\ee The model predicts $\La\sim 10^{10}$ GeV in the perturbative limit $h^R\sim 1$. Even relaxing the weak scale ratio as $u/v =$ 1000--0.001, the $B-L$ breaking scale is $\La=10^{16}$--$10^{4}$ GeV, respectively, which is beyond TeV scale, where the relevant new physics is governed by $w$. The second symmetry breaking scheme is most favored.       

Without loss of generality, consider the Yukawa couplings of charged leptons $h_{ab}^l$ to be flavor diagonal. Thus, the neutrino mixing is completely operated by $h_{ab}^R$, and this is also an important source for charged lepton flavor violating processes. Specially, the processes like $\mu \rightarrow3e $ happen at the tree level by the exchange of doubly-charged scalar ($\Xi^{\pm\pm}_{22}$), obtained by
\bea
\mathrm{Br}(\mu^- \rightarrow e^+e^-e^-) \simeq \fr{\Gamma(\mu^- \rightarrow e^+e^-e^-)}{\Gamma ( \mu^-\rightarrow e^- \nu_\mu \bar{\nu}_e)}=\fr{1}{G_F^2 m^4_{\Xi_{22}}}\lvert h^R_{e\mu}\lvert^2 \lvert h^R_{ee}\lvert^2,
\eea 
which is suppressed by the ${\Xi_{22}}$ mass, where the Fermi constant is $G_F = 1.16637 \times 10^{-5}\ \mathrm{GeV}^{-2}$. The present non-observation of the transition $\mu^- \rightarrow e^+e^-e^-$ bounds $\mathrm{Br}(\mu^- \rightarrow e^+e^-e^-) <10^{-12}$~\cite{data}, which translates to $h^R_{ee,e\mu}=10^{-3}$--1 for $m_{\Xi_{22}}\simeq \sqrt{-\la_{2\Xi}}\La=1$--1000 TeV, respectively. 

The processes like $\mu \rightarrow e \gamma$ are induced by one-loop corrections of two kinds. The first kind is mediated by the charged gauge bosons $W_{L,R}^\pm$ and $ Y_R^{\pm (q+1)}$ due to the neutrino and exotic-lepton mixings, respectively, which is very suppressed \cite{muegamma1, muegamma2}. The second kind is contributed by singly charged scalars and neutrinos, or doubly charged scalars and charged leptons $(\tau, \mu, e)$. Since the former contribution is similar to the first kind, the latter would dominate which leads to
\bea
\mathrm{Br}(\mu \rightarrow e \gamma) \simeq \fr{\al}{48\pi}\fr{25}{16}\fr{\rvert(h^{R\dag} h^R)_{12}\arrowvert ^2}{M^4_{\Xi_{22}}G_F^2},
\eea
where the fine structure constant is $\al = 1/128$. Taking the experimental bound $\mathrm{Br}(\mu \rightarrow e \gamma)<4.2 \times 10^{-13}$ \cite{data} leads to $m_{\Xi_{22}} = 1$--100 TeV for $|(h^{R\dagger} h^R)_{12}|=10^{-3}$--$10$, respectively. Comparing to the previous bound, this case translates to $h^R_{e\tau,\mu\tau}\simeq 0.03$--3.16.   

\section{\label{collisionb} Search for $\mathcal{Z}_1$ and $\mathcal{Z}'_1$ at colliders}

The new neutral gauge bosons have both couplings to leptons and quarks. The LHC can bound the quark couplings as well as the products of two coupling types, but not the lepton couplings only. Last one can be addressed by the lepton colliders.      

\subsection{LEPII}

The LEPII at CERN searched for new neutral gauge boson signals that mediate the processes such as $e^+ e^- \rightarrow (\mathcal{Z}_1, \mathcal{Z}_1^\prime) \rightarrow f \bar{f}$, where $f$ is ordinary fermion in the final state. From the neutral currents in \cite{2L3R}, we obtain effective interactions describing the processes,
 \bea
 \mathcal{L}_{\mathrm{eff}} &=& \fr{g_L^2}{c_W^2m^2_I}\left[ \bar{e} \ga^\mu (a_L^I(e) P_L+a_R^I (e) P_R)e\right] \left[\bar{f}(a_L^I(f) P_L+a_R^I (f) P_R)f \right]\crn
 &=& \fr{g^2_L}{c^2_W}\left(\fr{a^{\mathcal{Z}_1}_L(e)a^{\mathcal{Z}_1}_L(f)}{m^2_{\mathcal{Z}_1}}+\fr{a^{\mathcal{Z}'_1}_L(e)a^{\mathcal{Z}'_1}_L(f)}{m^2_{\mathcal{Z}'_1}}\right)(\bar{e}\ga^\mu P_L e)(\bar{f}\ga_\mu P_L f)\crn
 &&+(LR)+(RL)+(RR),
 \label{matr}\eea
where $I$ is summed over $\mathcal{Z}_1,\mathcal{Z}'_1$, and the chiral couplings $a^I_{L,R}(f)=[g^I_V(f)\pm g^I_A(f)]/2$ can be extracted from \cite{2L3R}.

Concretely, the LEPII searched for such chiral interactions and gave the bounds on respective effective couplings. The most relevant one includes left-handed fermions for $f=\mu$, yielding \cite{LEP}
\bea
\fr{g^2_L}{4  c_W^2} \fr{1}{t_R^2+\beta^2 t_X^2} \left(\fr{(s_\epsilon s_W+ c_\epsilon  c_W \beta  t_X)^2}{m^2_{\mathcal{Z}_1^2}} + \fr{(c_\epsilon s_W-c_W s_\epsilon \beta t_X)^2}{m^2_{\mathcal{Z}_1^\prime}}  \right) < \fr{1}{ (6\ \mathrm{TeV})^2}.
\label{lep1}\eea 
As determined in the neutrino mass section, we have $\La\gg w$, thus $m_{\mathcal{Z}'_1}\gg m_{\mathcal{Z}_1}$. Only $\mathcal{Z}_1$ contributes, leading to the $m_{\mathcal{Z}_1}$ bound as 
\be m_{\mathcal{Z}_1}>\fr{3g_L}{c_W}\fr{s_\epsilon s_W+ c_\epsilon  c_W \beta  t_X}{\sqrt{t_R^2+\beta^2 t_X^2}}\ \mathrm{TeV}.\ee 
The mixing angle $\epsilon$ is finite, depending only on the gauge couplings and $\beta$, due to $\La\gg w$ \cite{2L3R}. Taking $t_R=1$ and $t_X=s_W/\sqrt{1-(2+\beta^2)s^2_W}$, we get $m_{\mathcal{Z}_1}>\mathcal{O}(1)$ TeV.   

\subsection{LHC}

We consider only the $\mathcal{Z}_1$ processes at the LHC since $\mathcal{Z}'_1$ is superheavy and does not contribute. Because $\mathcal{Z}_1$ directly couples to the standard model quarks, it may be produced at the LHC by $s$-channel and then decays into high energy lepton and/or jet pairs. Especially, the leptonic productions are very attractive for studying heavy neutral gauge bosons with unsuppressed couplings to leptons \cite{ALA}. In the narrow width approximation, the cross-section for producing a $\mathcal{Z}_1$ boson at the LHC and then decaying into a $f \bar{f}$ final state takes the form \cite{ALA1}
\bea
\sigma(pp \rightarrow \mathcal{Z}_1 \rightarrow f \bar{f})= \left[\fr{1}{3} \sum_{q=u,d}\left(\fr{dL_{q \bar{q}}}{dm^2_{\mathcal{Z}_1}} \right) \hat{\sigma}(q\bar{q} \rightarrow \mathcal{Z}_1)\right ] \times \mathrm{Br}(\mathcal{Z}_1 \rightarrow f \bar{f}).
\eea
In what follows, we consider the parton luminosity $dL_{q \bar{q}}/dm^2_{\mathcal{Z}_1}$ at the LHC for $\sqrt{s}=13$ TeV which can be directly obtained from the first reference of \cite{ALA2}. 
\begin{figure}[H]
	\centering
	\begin{tabular}{ccc}
		\includegraphics[width=12cm]{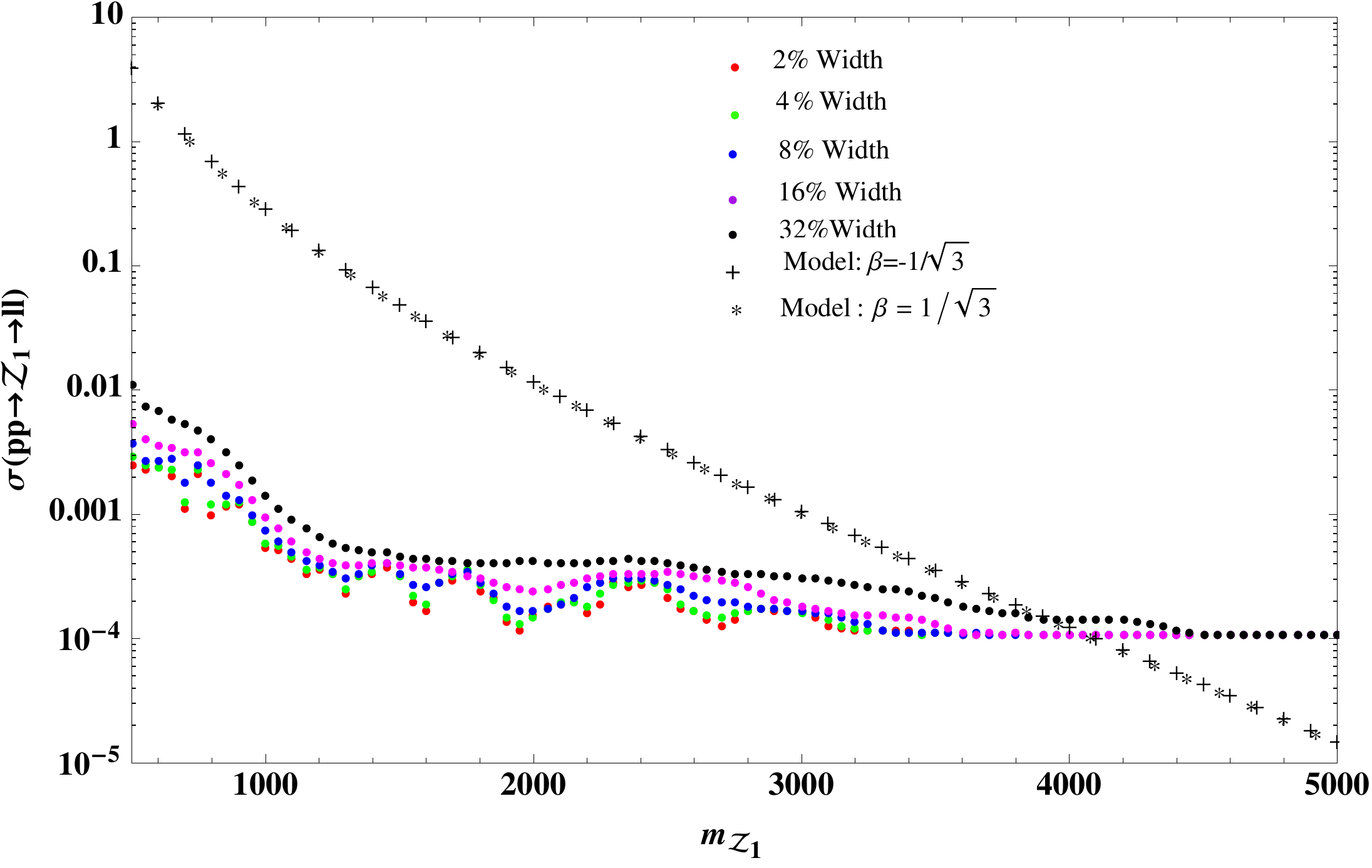}  
	\end{tabular}
	\caption{ \label{hinhLHC} The cross-section $\sigma(pp\rightarrow \mathcal{Z}_1\rightarrow l\bar{l})$ [pb] as a function of $m_{\mathcal{Z}_1}$ [GeV], where the points are the observed limits according to the different widths extracted at the resonance mass in the dilepton final state using 36.1 fb$^{-1}$ of proton-proton collision data at $\sqrt{s}= 13$ TeV with ATLAS detector \cite{Alast2017}. The star and plus lines are the theoretical predictions for $\beta=\pm 1/\sqrt{3}$, respectively.}
\end{figure} 

In Fig. \ref{hinhLHC}, we show the cross-section for the tree-level process 
$pp \rightarrow \mathcal{Z}_1 \rightarrow l \bar{l}$, where $l$ is either electron or muon which has the same coupling to $\mathcal{Z}_1$, for two versions $\beta = \pm 1/\sqrt{3}$. Both the theoretical predictions are nearly close, weakly separated by different $\beta$ signs. The experimental search uses 36.1 fb$^{-1}$ of proton-proton collision data, collected at $\sqrt{s}=13$ TeV by the ATLAS experiment \cite{Alast2017}, giving a negative signal for new high-mass phenomena in the 
dilepton final state. It is converted into the lower limit on the $\mathcal{Z}_1$ mass, $m_{\mathcal{Z}_1} > 4$ TeV, for models
with $\beta=\pm 1/\sqrt{3}$. 

\section{\label{VCT} Dark matter phenomenology}

In this section, we study the dark matter observables corresponding to the two dark matter versions for $q=0$ and $q=-1$ as obtained before.

\subsection{The 3-2-3-1 model with $q=0$} 

In this model, the dark matter candidates are $E_{1,2,3}^0, H_6^0, H_7^0, X_{R}^0$. Recall that the states $H^0_{6,7}$ and $X^0_R$ have the masses proportional to $\La$ scale, while the neutral fermions $E^0_{1,2,3}$ have the masses proportional to $w$ scale. Since $\La\gg w$, the LWP is naturally taken as a light combination of $E^0_{1,2,3}$, called $E^0$. However, if one finetuns the self-scalar couplings or $g_R$, the LWP may also be a scalar or a vector.    
Depending on the parameter space, we consider three cases.
   
\subsubsection{Fermion dark matter} 

 Supposing that $E_1$ is the lightest state among all the $W$-particles, it is stabilized responsible for dark matter due to $W$-parity conservation and kinetic suppression. $E_1$ directly couples to the normal leptons $\nu,l$ via the new gauge bosons $X_{R}^{0,0*}, Y_{R}^{\pm 1}$, respectively, and it also has the neutral currents with $\mathcal{Z}_1, \mathcal{Z}_1^\prime$. Denote the remaining lepton flavors by $\nu_\al, l_\al$. $E_1$ dominantly annihilates into the standard model particles as 
 \bea
 E_1 E_1^c \rightarrow \nu\nu^c, l^-l^+, \nu_\al \nu^c_\al, l_\al^- l_\al^+, qq^c, ZH_1,
 \label{huy1}\eea where the first two productions have both $t$-channel by respective $X_R,Y_R$ and $s$-channel by $\mathcal{Z}_1, \mathcal{Z}_1^\prime$, while the remainders have only the $s$-channel. There may exist some contributions from the new scalar portals, but they are small and neglected. There is no standard model Higgs or $Z$ portal.  
 
 The neutral gauge bosons $\mathcal{Z}_1, \mathcal{Z}_1^\prime$ mix via a finite angle, $\epsilon$, and their interactions can be interchanged by replacing ($c_\epsilon \rightarrow -s_\epsilon,s_\epsilon \rightarrow c_\epsilon$), respectively. Therefore, they play a similar role in the dark matter annihilation channels given in (\ref{huy1}). However, we stress that the contributions of $\mathcal{Z}_1, \mathcal{Z}_1^\prime$ to the dark matter annihilation processes are proportional to $\fr{1}{4m^2_{E_1}-m^2_{\mathcal{Z}_1}}$, $\fr{1}{4m^2_{E_1}-m^2_{\mathcal{Z}_1^\prime}}$, respectively. Due to the condition $\La \gg w$, or correspondingly $m^2_{\mathcal{Z}_1^\prime} \gg m^2_{\mathcal{Z}_1}$, the field $\mathcal{Z}_1$ is active that dominantly sets the dark matter observables. Also in this limit, $m^2_{X_R} \gg m^2_{Y_R}$, only the charged gauge boson $Y_R$ contributes to the $t$-channel, but radically smaller than those of $\mathcal{Z}_1$. The dark matter is stabilized if $m_{E_1} < m_{Y_R}$. We also take $g_L=g_R$ for calculations.   

\begin{figure}[H]
 	\centering
 	\begin{tabular}{ccc}
 		\includegraphics[width=5cm]{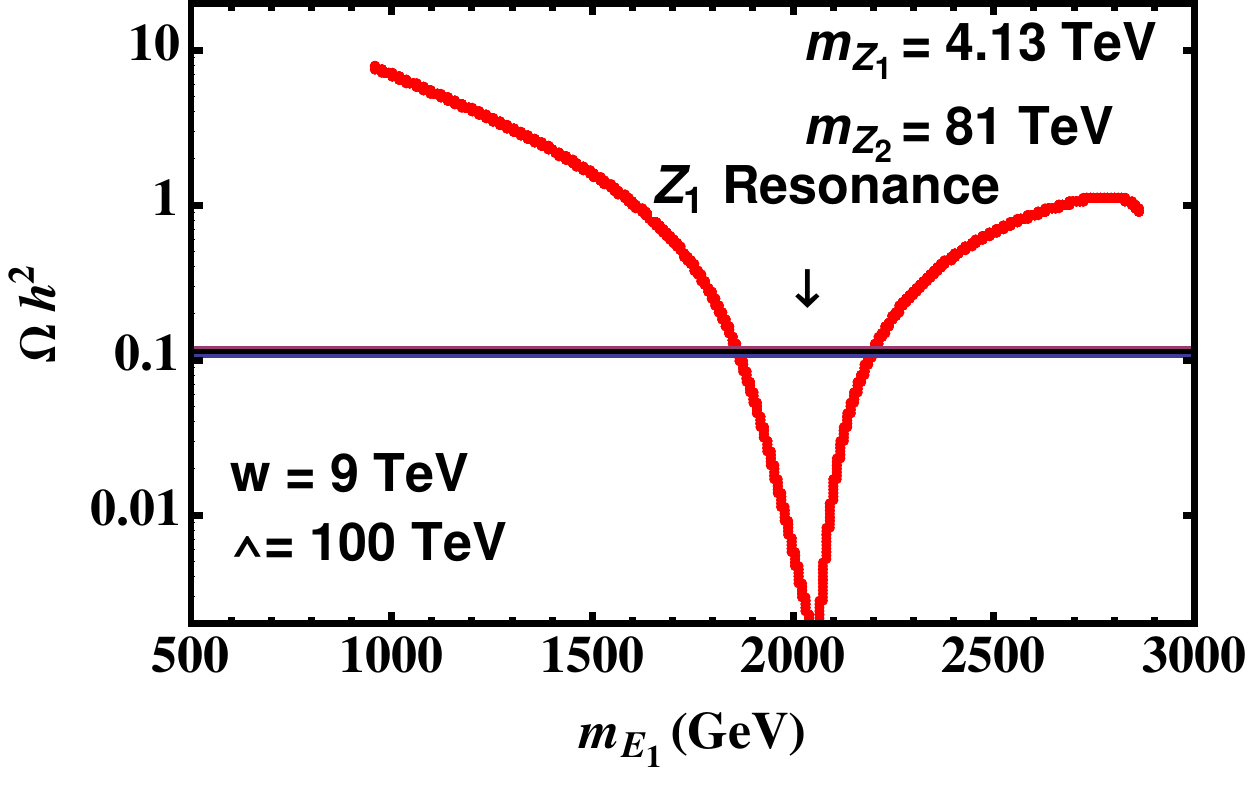} &
 		\includegraphics[width=5cm]{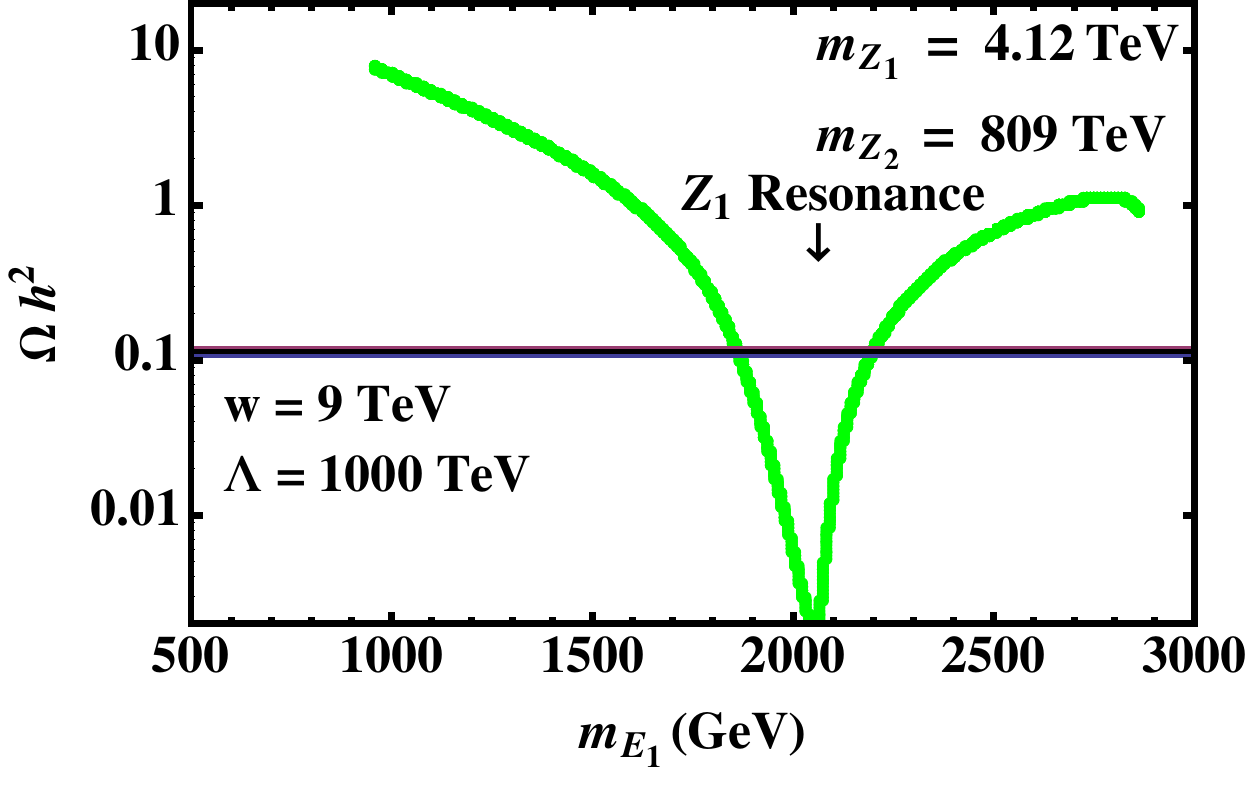} &
 		\includegraphics[width=5cm]{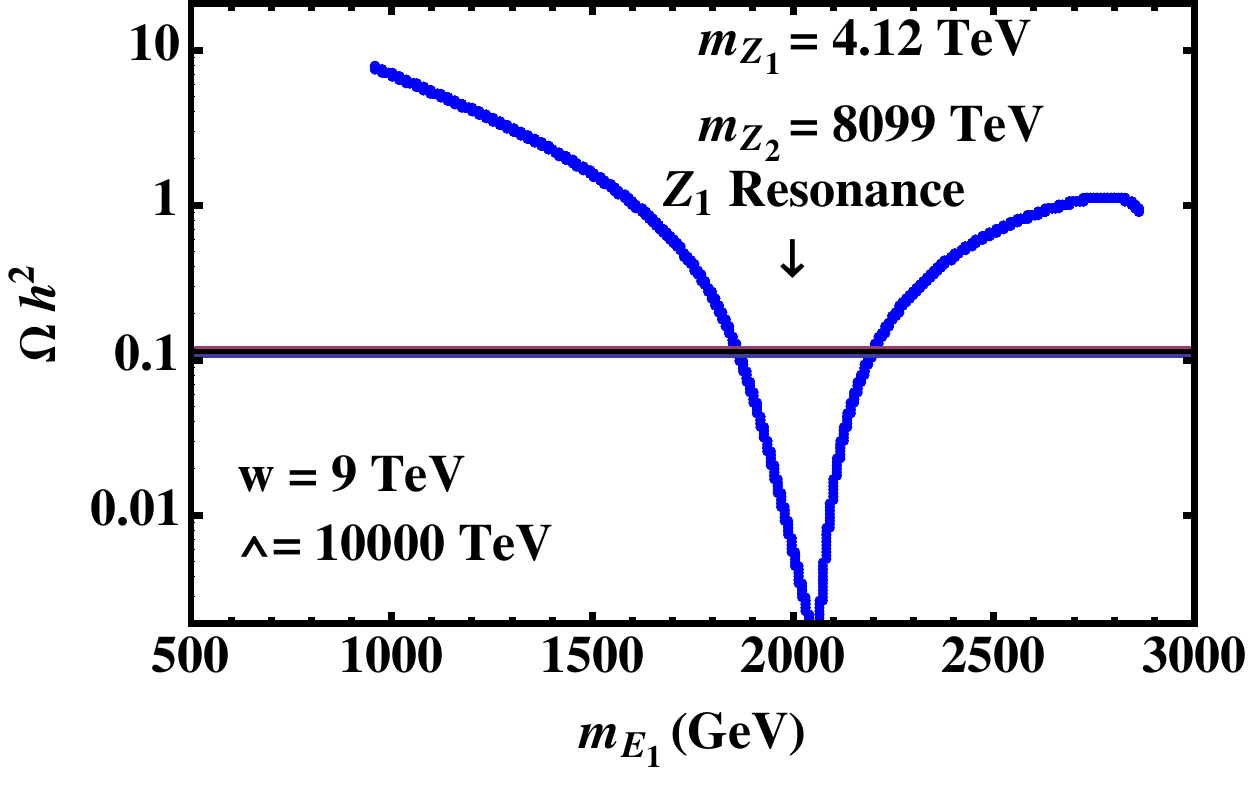}  
 	\end{tabular}
 	\caption{The relic density of the fermion candidate as a function of
 		its mass in the limit $\La \gg w$, where we label $Z_1\equiv \mathcal{Z}_1$ and $Z_2\equiv \mathcal{Z}'_1$ which should not be confused.} \label{hinh1}
 \end{figure}

In Fig. \ref{hinh1} we display the dark matter relic density as a function of its mass. The panels from left to right correspond to the selections of the $\mathcal{Z}_1^\prime$ mass as $81, 809, 8099$ TeV, respectively. It is clear that the relic density is almost unchanged when $m_{\mathcal{Z}_1^\prime}$ changes. The stabilization of dark matter yields only a $\mathcal{Z}_1$ resonance regime. For instance, $w=9$ TeV, the dark matter mass region is $1.85<m_{E_1}< 2.15$ TeV, given that it provides the correct abundance. 
 
\begin{figure}[H]
		\centering
		\begin{tabular}{ccc}
			\includegraphics[width=5cm]{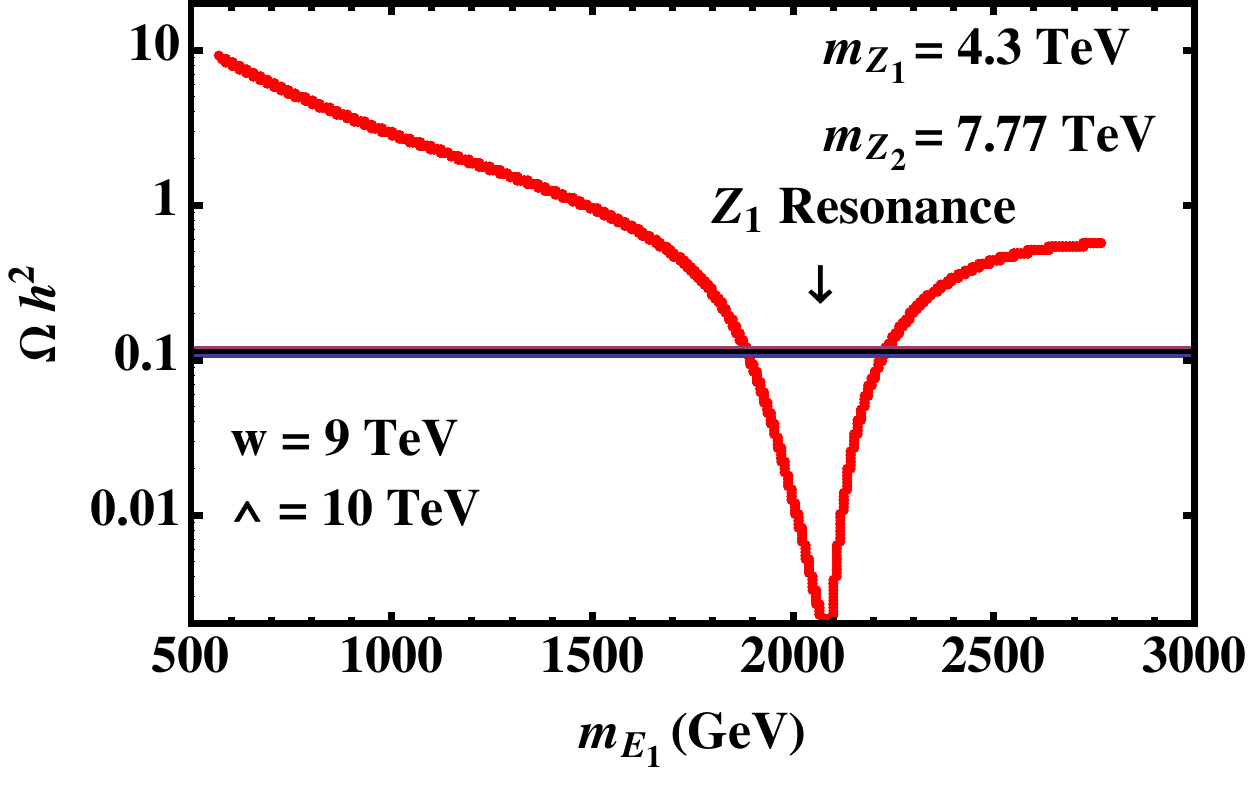} &
			\includegraphics[width=5cm]{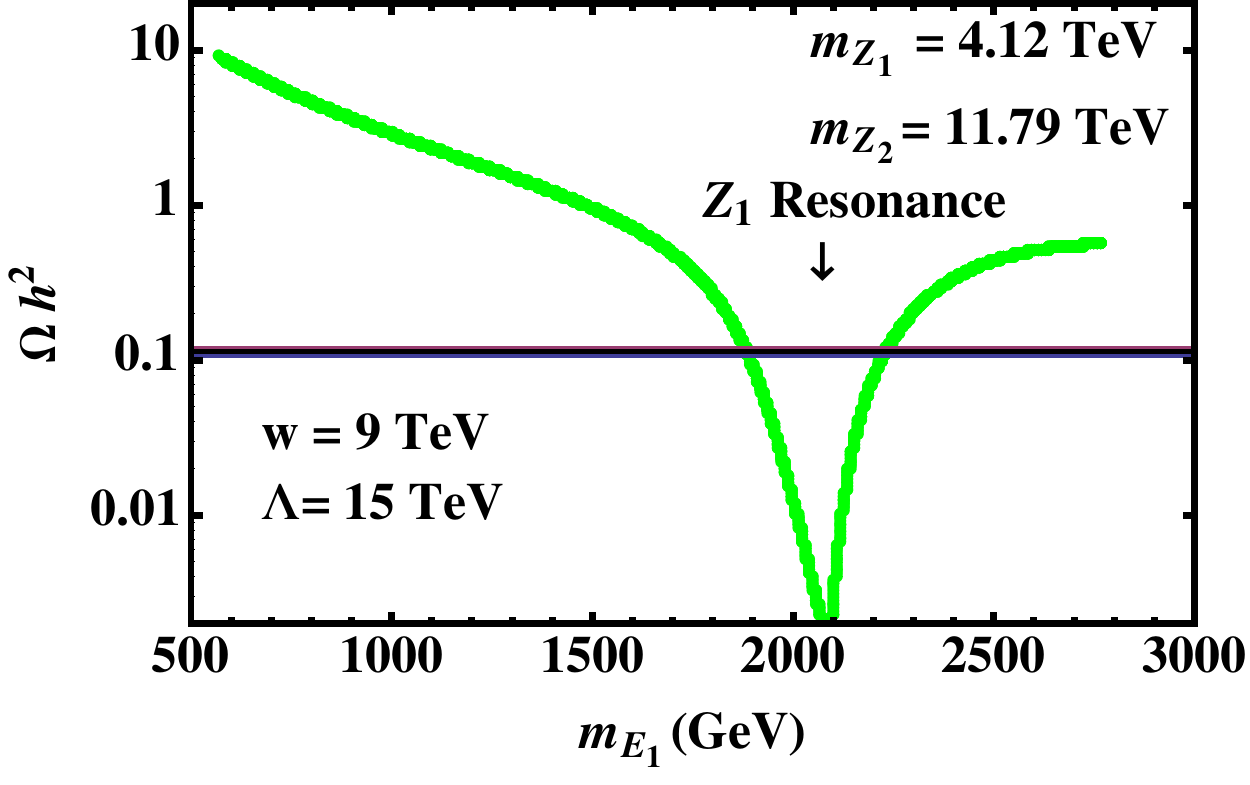} &
			\includegraphics[width=5cm]{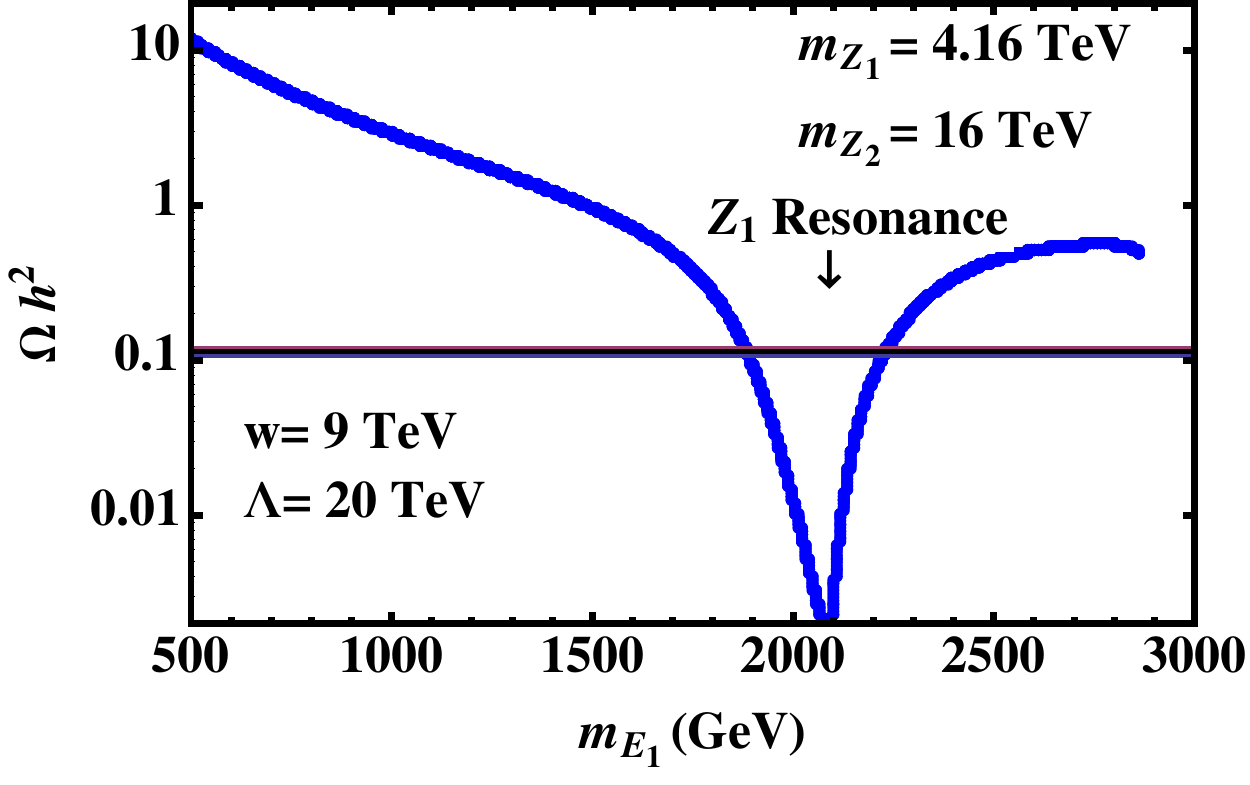}  
		\end{tabular}
		\caption{The relic density of the fermion candidate as a function of
     its mass in the limit $ \La \gtrsim w$, where $Z_1\equiv \mathcal{Z}_1$ and $Z_2\equiv \mathcal{Z}'_1$.}\label{hinh2}
	\end{figure}

If one relaxes the constraint from the neutrino mass generation by setting $\La\gtrsim w$, the $\mathcal{Z}_1^\prime$ contribution may become significant. Note that in this case the mixing angle $\epsilon$ is also finite and the $\mathcal{Z}_1$ and $\mathcal{Z}_1^\prime$ couplings to fermions are equivalent. However, since $m_{\mathcal{Z}_1} \lesssim m_{\mathcal{Z}_1^\prime}$ and to protect $m_{E_1}<m_{Y_R}$, only the $\mathcal{Z}_1$ resonance regime exists, as depicted in Fig. \ref{hinh2}.         

\begin{figure}[H]
		\centering
		\begin{tabular}{ccc}
			\includegraphics[width=5cm]{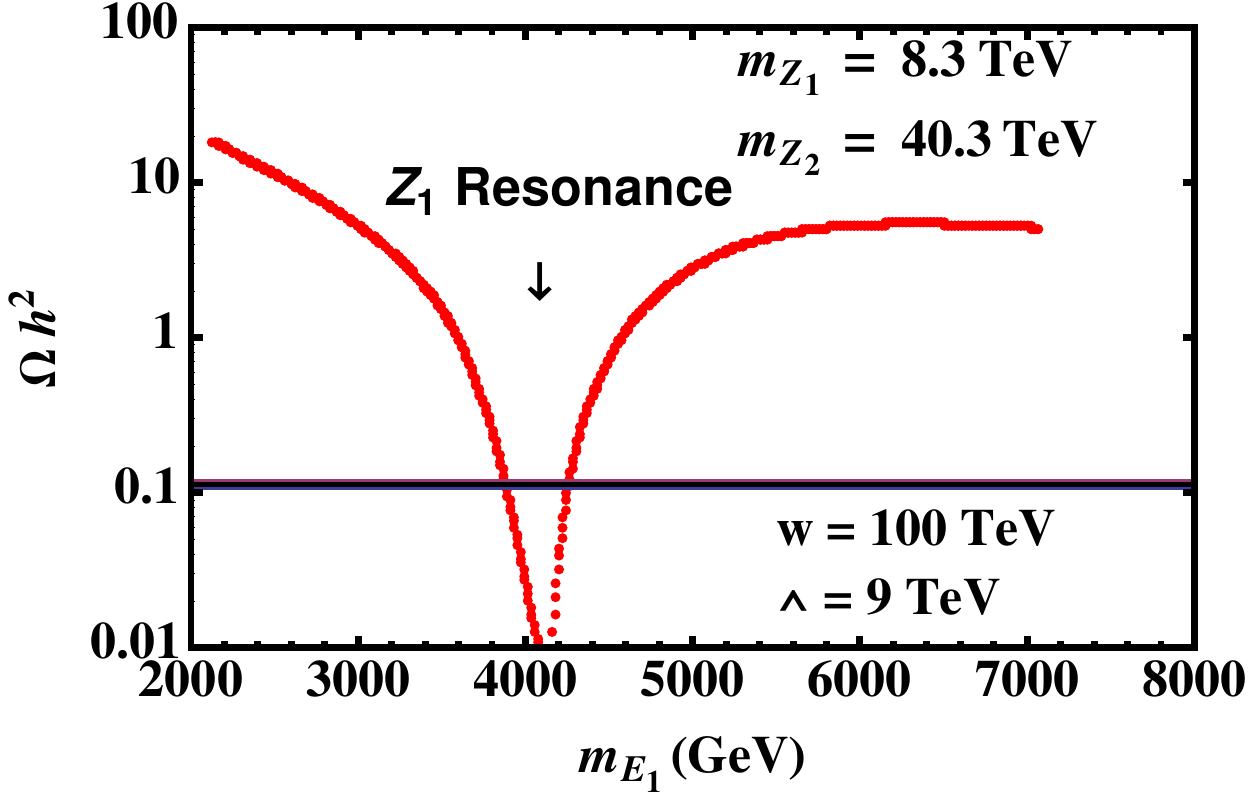} &
			\includegraphics[width=5cm]{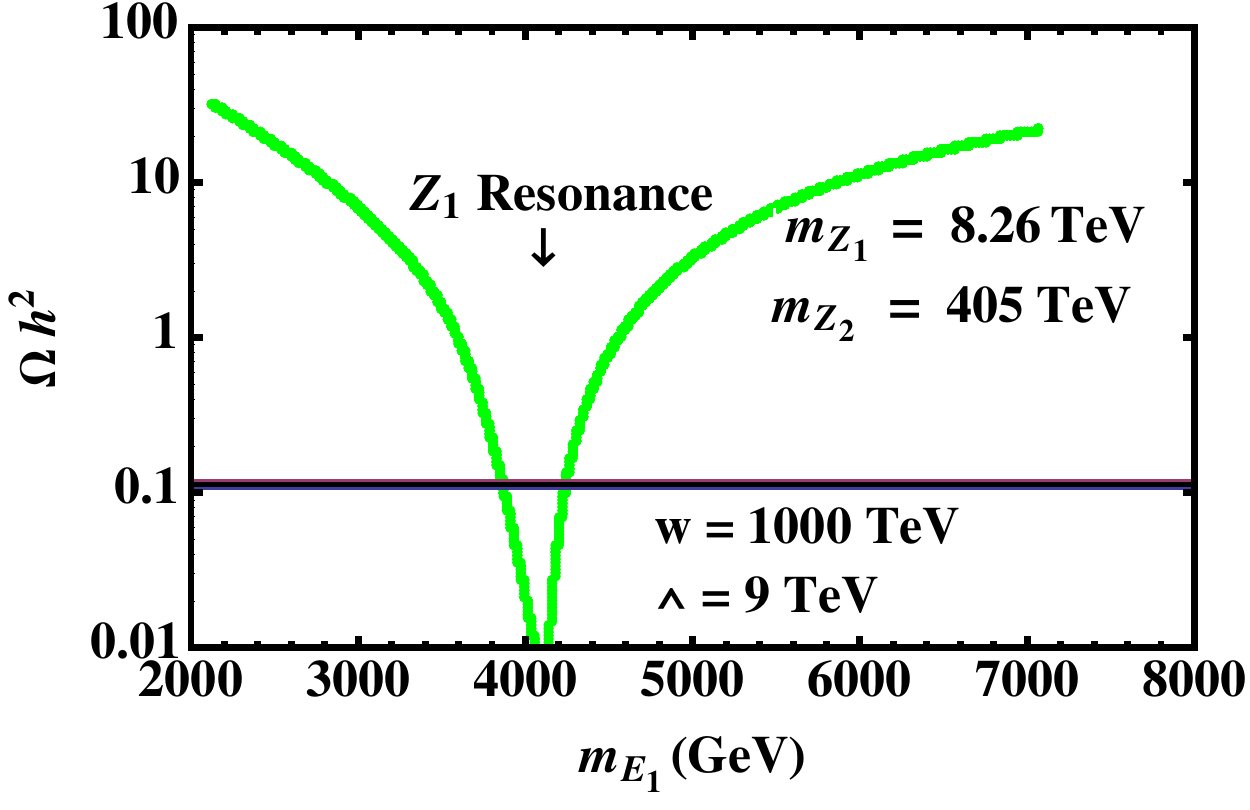} &
			\includegraphics[width=5cm]{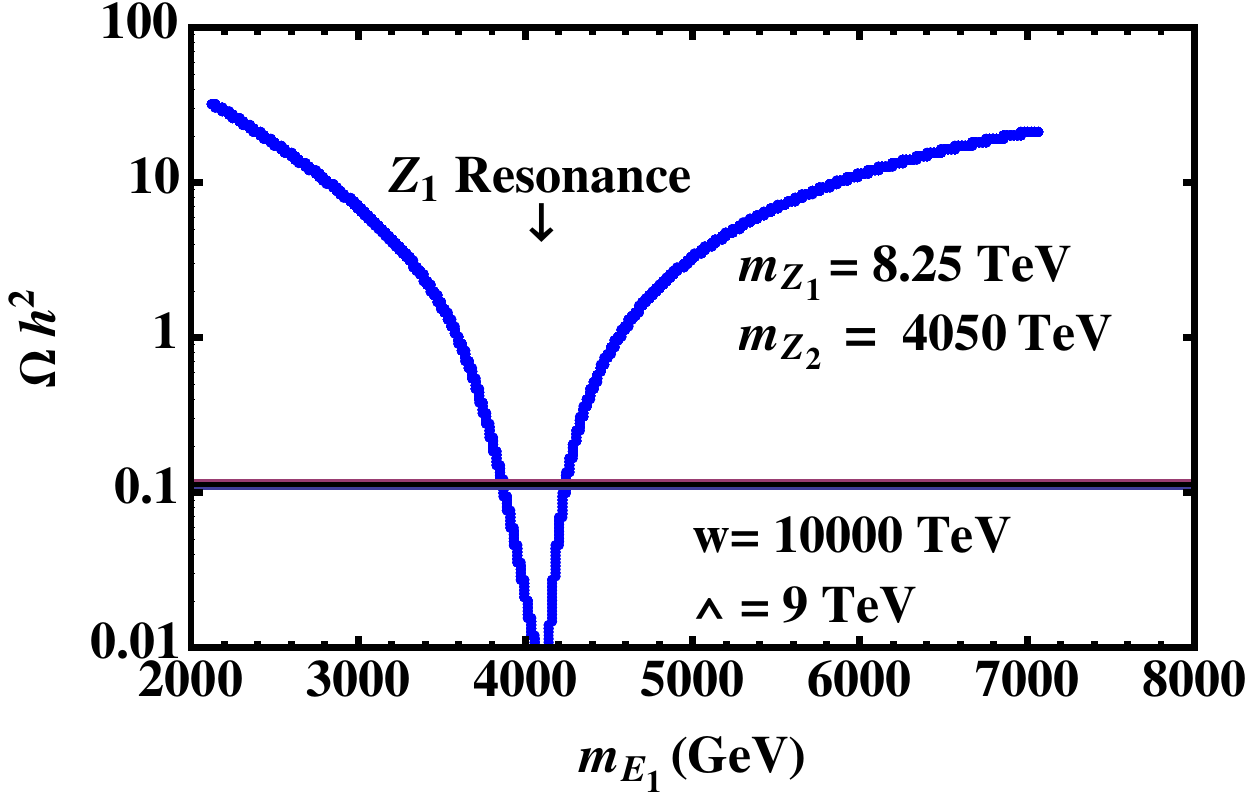}  
		\end{tabular}
		\caption{The relic density of the fermion candidate as a function of its mass, $m_{E^0}$, in the limit $w \gg \La$, where $Z_1\equiv \mathcal{Z}_1$ and $Z_2\equiv \mathcal{Z}'_1$.}\label{hinh3}
	\end{figure}

Even if one considers $w\gg\La$. In this case, the mixing effect of $\mathcal{Z}_1$ and $\mathcal{Z}_1^\prime$ is very small, but still $m_{\mathcal{Z}_1}\ll m_{\mathcal{Z}_1^\prime}$. Therefore, the $\mathcal{Z}_1$ gauge boson dominates the annihilation as given before. Similar to the case in Fig. \ref{hinh1}, the resonance regime for $\mathcal{Z}_1$ is not sensitive to the change of the large VEV as plotted in Fig. \ref{hinh3}. 
	
In short, $\mathcal{Z}_1$ always governs and sets the dark matter density in all cases, i.e. it is just active portal, provided $E_1$ is stabilized as leading to only a resonance region for $\mathcal{Z}_1$.  

\begin{figure}[H]
	\centering
	\includegraphics[scale=0.6]{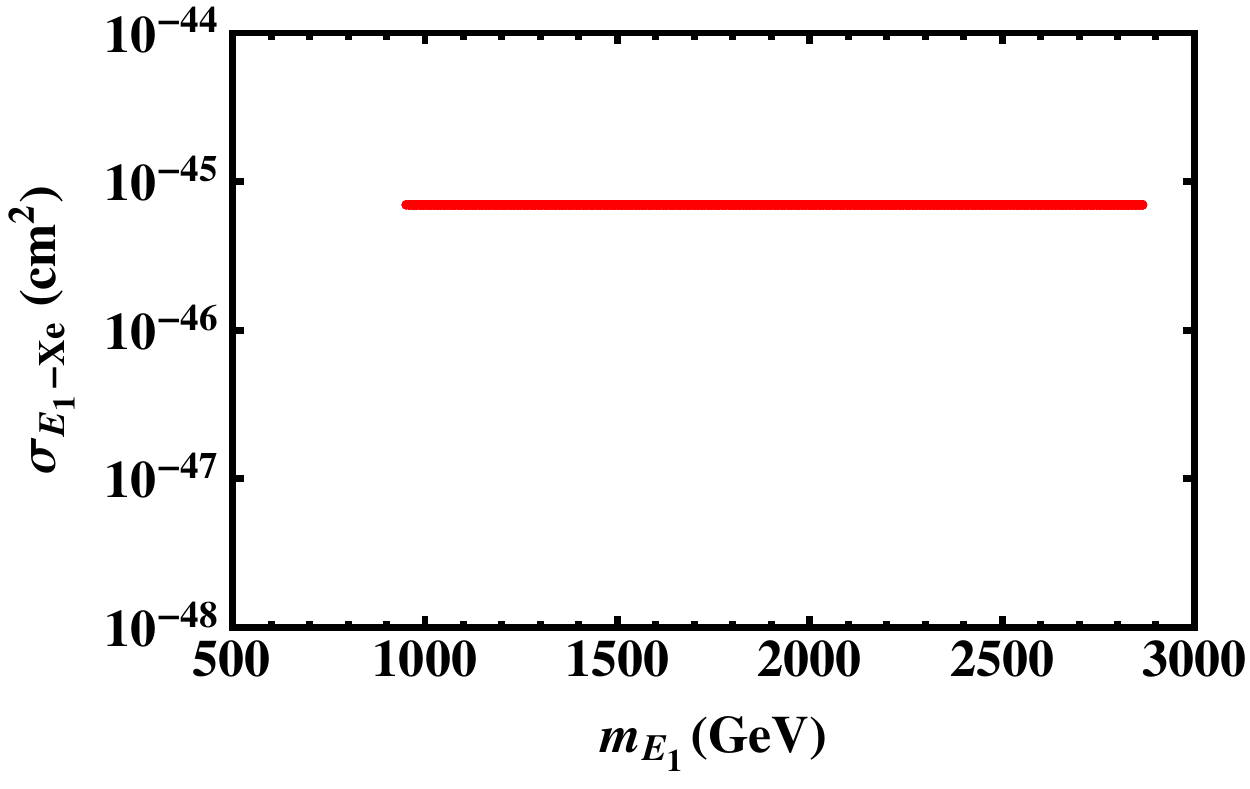}
	\includegraphics[scale=0.6]{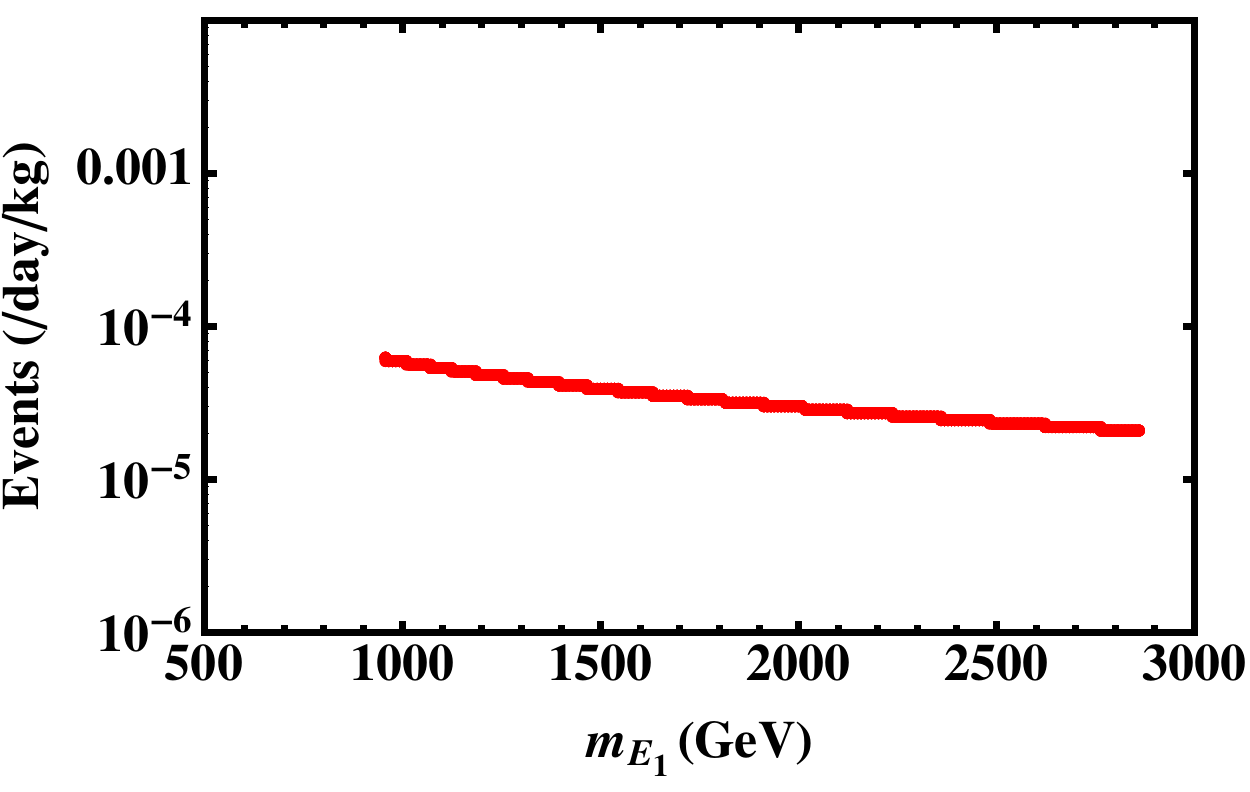}
	\caption{The scattering cross-section (left-panel) and the total number of events/day/kg (right-panel) as functions of fermion dark matter mass. \label{hinh4}}
\end{figure}

The direct detection experiments measure the recoil energy deposited by the scattering of dark matter with the nuclei. This scattering is due to the interactions of dark matter with quarks confined in nucleons. The scattering amplitude comes from t-channels via the exchanges of $\mathcal{Z}_1, \mathcal{Z}_1^\prime$ bosons. There exist   
both spin-independent and spin-dependent interactions, but for heavy nuclei, the cross-section is enhanced by the spin-independent interactions due to the factor $A^2$. Using micrOMEGAs 4.3.5, we get the cross-section for $E_1$-nucleon elastic scattering and the total number of events/day/kg for detector $Xe$. Fig. \ref{hinh4} shows that the predicted results are consistent with the XENON1T experiment \cite{Xenon} since the dark matter mass is in the TeV scale. 

Let us remind the reader that when the coupling strength of dark matter and normal matter is similar to the electroweak couplings like our model, the dark matter was generally thought to overpopulate the universe for heavy candidates at TeV scale. And, this was the reason why the WIMP was often interpreted to have a mass at the weak scale, $m_{\mathrm{WIMP}}\sim 150$ GeV, because $\Om h^2\simeq 0.1\ \mathrm{pb}/\langle \sigma v \rangle$ and $\langle \sigma v \rangle \simeq \al^2/(150\ \mathrm{GeV})^2\simeq 1\ \mathrm{pb}$ recovers the observed relic density $\Om h^2\simeq 0.11$~\cite{data}. However, the solution changes when dark matter interacts with normal matter via a s-channel heavy portal. Indeed, our fermion dark matter dominantly annihilates into the standard model particles via s-channel by $\mathcal{Z}_1$. The cross-section is proportional to the squared $\mathcal{Z}_1$ propagator, $\langle \sigma v \rangle\sim 1/{(4m^2_{E_1}-m^2_{\mathcal{Z}_1})^2}$, given in the center-of-mass frame. Thus, the relic density is $\Om h^2\sim 1/{\langle \sigma v \rangle }\sim (4m^2_{E_1}-m^2_{\mathcal{Z}_1})^2$. Hence, we have a resonance at $m_{E_1}=\fr 1 2 m_{\mathcal{Z}_1}$, at which the large relic density rapidly decreases to zero. Due to the nature of a resonance, the right relic density is only a narrow funnel, with the width at funnel top proportional to $2\fr{v^2_{\mathrm{weak}}}{m_{\mathcal{Z}_1}}\sim 30$ GeV, and we can say that the resonance sets the dark matter observables. A consequence of this analysis is that the vector candidates are ruled out, since they have additionally contact interactions to $W,Z$ that govern the relic density, as shown below. On the other hand, of course such heavy dark matter would not be restricted by the direct or indirect detections. However, we would like to include Fig. \ref{hinh4} for concreteness and the fact that the $\mathcal{Z}_1$ mass limit may be raised if the future search is continuously negative. To conclude, only the points around $m_{E_1}=\fr 1 2 m_{\mathcal{Z}_2}\simeq 2$ TeV in Fig. \ref{hinh4} respect the relic density bounds for $m_{\mathcal{Z}_1}\simeq 4$ TeV, appropriate to the above LHC dilepton search.  

\subsubsection{Scalar dark matter\label{SDM}}

In the limit $w, \La \gg u, v$, the scalar $H_6$ transforms as a $SU(2)_L$ doublet while $H_7$ is a $SU(2)_L$ singlet. If $H_6$ is the LWP, it has the properties of dark matter as in the inert doublet model~\cite{inertdoublet}. The field $H_6$ can annihilate into $W^+W^-, ZZ, H_1H_1$ and $\bar{f}f$ since its mass is beyond the weak scale. Generalizing the result from M. Cirelli {\it et al.} in \cite{ad2}, the annihilation cross-section is given by $\langle \sigma v \rangle \simeq \left(\fr{\al}{150\ \mathrm{GeV}}\right)^2\left[\left(\fr{600\ \mathrm{GeV}}{m_{H_6}}\right)^2+\left(\fr{x\times 1.354\ \mathrm{TeV}}{m_{H_6}}\right)^2\right]$, where $x\equiv \sqrt{\la^2_{1S}+\la^2_{2S}}$, $(\al/150\ \mathrm{GeV})^2\simeq 1$~pb aforementioned, and the first and second terms in the brackets come from the standard model gauge and Higgs portal interactions of $H_6$, respectively. From the Higgs mass constraint~(\ref{neutralHiggs}), $\la_{1S,2S}$ are proportional to the standard model Higgs self-coupling, thus $x\sim \la_{\mathrm{SM}}\simeq 0.127$. Hence, in the most area of the parameter space between the weak and new physics scales,
the annihilation through the gauge portal to $W^+W^-$ and $ZZ$ is so effective (i.e., dominant), and that it derives the thermal abundance equally to or
below the measured value for $m_{H_6}<600$ GeV. Above this value, the relic density is overpopulated. However, when $m_{H_6}$ is large, the scalar dark matter can (co)annihilate into the new normal particles of the 3-2-3-1 model via the new gauge and Higgs portals similarly to the 3-3-1 model \cite{ad5}, and this can reduce the abundance of dark matter to the observed value, in agreement with the experimental data \cite{darkrv}. Unfortunately, in our case, the scalar doublet dark matter $H_6$ may scatter off nuclei via $t$-channel $Z$ exchange, which induces a large cross-section and is already ruled out by the direct detection experiments (see R. Barbieri {\it et al.} in \cite{inertdoublet} for details). Moreover, due to $W$-parity conservation, the real and imaginary parts of $H_6$ always have degenerate masses; therefore, there is no way to suppress such channel, unlike the case of the inert doublet model. So, this candidate is not further discussed.      

\begin{figure}[h]
	\centering
	\includegraphics[scale=0.9]{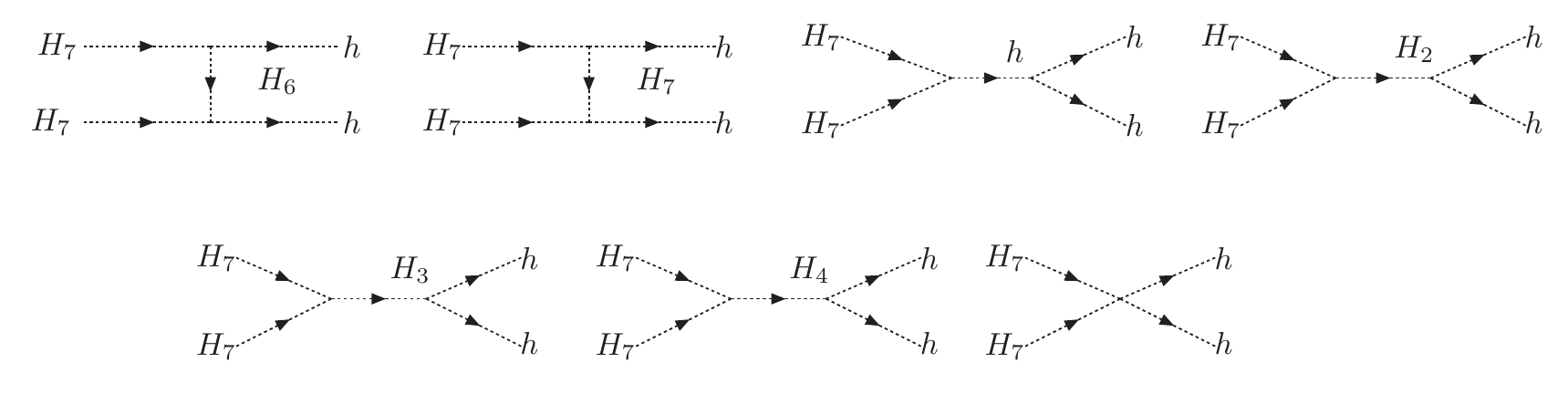}
	\caption{Diagrams that describe the annihilation $H^*_7 H_7 \rightarrow H_1 H_1$ via the Higgs portals, where and in the text we sometimes denote $h\equiv H_1$ for brevity.}
	\label{FMH7}\end{figure}

Let us assume the scalar $H_7$ as a dark matter candidate, which is now the LWP and leading to a 
condition $\la_3< \fr{g_R^2 w^2}{w^2+ 2\La^2}$ due to $m_{H_7}<m_{Y_R}$. Additionally, the new fermion Yukawa couplings are chosen to be $h^{E,J}\geq g_R/\sqrt{2}$, so that $m_{E,J}\geq m_{Y_R}$ \cite{2L3R}. Because $\La \gg w$, the other $W$-scalars have masses proportional to $\La$, that are heavier than $H_7$, as expected. Since $H_7$ is a singlet of the $SU(2)_L$ group, it has only the Higgs ($H_{1,2,3,4,6,7}$), new gauge, and new fermion portals. The annihilation products can be the standard model Higgs, $W,Z$, top quark, and new particles. The most interesting case is to impose the parameter space so that the Higgs portal governs the dark matter observables. For this aim, we derive 
\bea V_{\mathrm{scalar}} &\supset& \fr 1 2 H^*_7 H_7 H^2_1 \left(\la_4+\fr{u^2}{u^2+v^2}\la_1\right)+H^*_7 H_7 H_1\sqrt{u^2+v^2}\left(\la_4+\fr{u^2}{u^2+v^2}\la_1\right)\crn
&&-H^*_7 H_7 H_2\fr{\la_1 u v}{\sqrt{u^2+v^2}}+H^*_7 H_7 H_3 w \left[2\la_\phi-\fr{\la^2_5}{2(\la_{1\Xi}+\la_{2\Xi})}\right]+H^*_7 H_7 H_4 \la_5 \La\crn
&&+\cdots \eea Here, note that $H_7\simeq \phi_1$, $H_3\simeq S_3$, $H_4\simeq S_4$, the $S_3$-$S_4$ mixing angle $\varphi\simeq \la_5 w/2(\la_{1\Xi}+\la_{2\Xi})\La\ll 1$, $m^2_{H_3}\simeq 2[\la_\phi-\la^2_5/4(\la_{1\Xi}+\la_{2\Xi})]w^2$, $m^2_{H_4}\simeq 2(\la_{1\Xi}+\la_{2\Xi})\La^2$, and $m^2_{H_2}=-\fr{\la_2(u^2+v^2)}{2(u^2-v^2)}\La^2$. Hence, the contact interaction (first term and fifth term after integrating $H_4$ out) and the $H_3$ portal (fourth term) set the relic density, while the $H_1$ portal (second term) sets the direct detection cross-section, provided that $\bar{\la}\equiv \la_4+\fr{u^2}{u^2+v^2}\la_1\sim 1$, $\bar{\la}'\equiv 2\la_\phi-\fr{\la^2_5}{2(\la_{1\Xi}+\la_{2\Xi})}\sim 1$, and $\la_5\sim 1$ are larger than $g_{L,R}$. The above analysis can fully demonstrated by the Feynman diagrams in Fig.~\ref{FMH7}, where the $t$-channels by $H_{6,7}$ are also included. The annihilation channels $H^*_7 H_7 \rightarrow H_1 H_1$ via the three graphs of the second row play a major role in determining the abundance, whereas the ones with $s$-channel by $H_{1,2}$ are suppressed by small couplings and heavy mediators \cite{3311dm}. Note that the trilinear Higgs couplings, $H_7H_7 h$ and $H_7 H_6 h$, depend on the $f$ parameter, $f= -\fr{\la_{2S}uv}{\sqrt{2}w}-\fr{\la_2uv\La^2}{2\sqrt{2}(u^2-v^2)w}\sim \la_2 \La^2/w$~\cite{2L3R}. If $\la_2$ is sizable, i.e. $f\sim \La$, the corresponding diagrams mediated by $H_6, H_7$ overwhelm annihilation processes, leading to almost vanishing relic density ($H_7$ annihilates completely, before freezeout). We concern a small coupling, say $\la_2\sim \la_3$ or $f\sim w$ (and, of course, $m_{H_6}\gtrsim m_{H_7}$), the relevant $t$-channel diagrams negligibly contribute, since their amplitudes are proportional to $f^2/m^2_{H_{6,7}}\sim g^2_R$, provided that $\la_2\sim \la_3\sim g^2_R w^2/\La^2$, which are manifestly suppressed due to the conditions, $m_{H_7}<m_{Y_R}$ and $g_R < \bar{\la},\bar{\la}', \la_5$, as mentioned.  
	
\begin{figure}[h]
	\centering
	\includegraphics[scale=0.8]{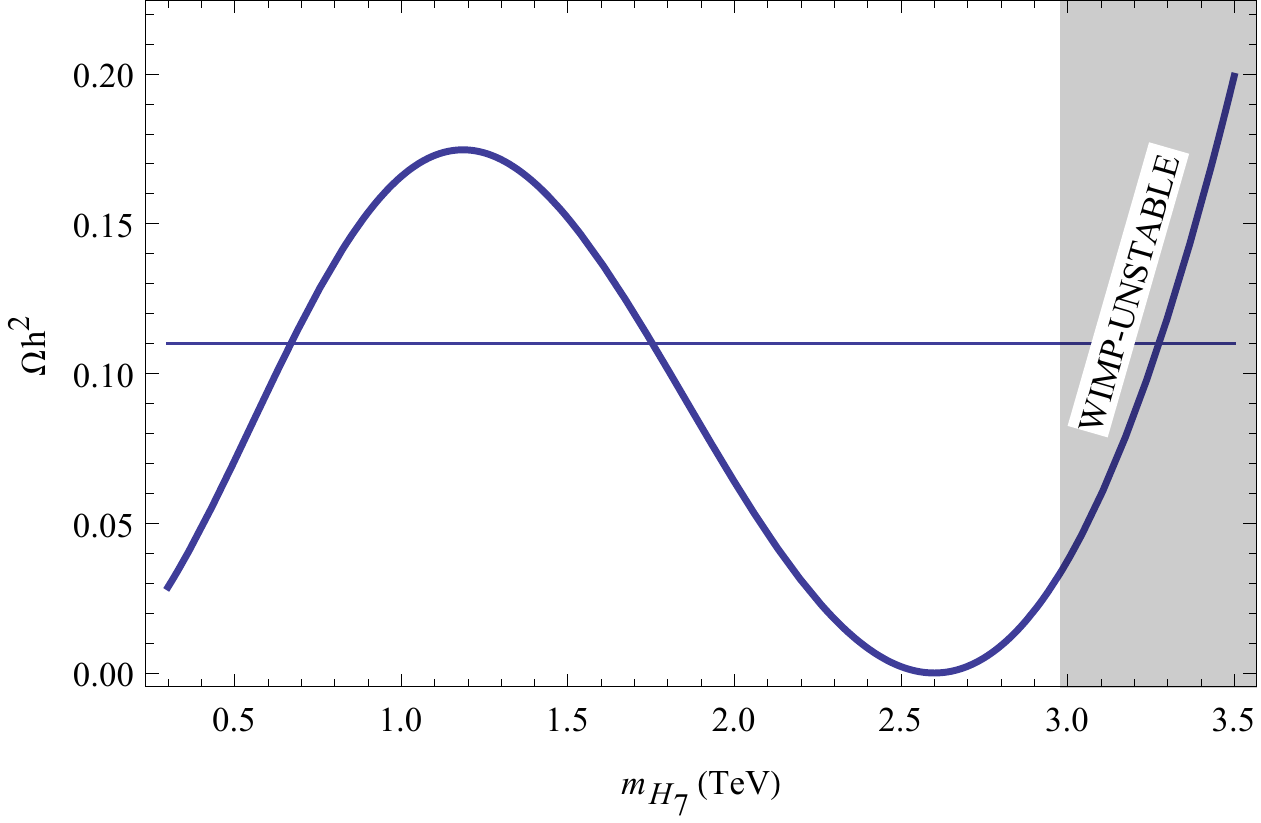}
	\caption{The relic density depicted as a function of the scalar $H_7$ mass.}
	\label{scalarH7}\end{figure}

That said, the dark matter annihilation amplitude is governed by the contact and $H_{3,4}$ portal interactions, given by 
\be M=-\bar{\la}+\fr{\la_5\la_6}{2(\la_{1\Xi}+\la_{2\Xi})}-\la'\fr{m_{H_3}^2}{4m^2_{H_7}-m^2_{H_3}},\ee where the $H_3 hh$ and $H_4hh$ couplings take the form $\mathcal{L}\supset -\fr 1 2 \la' w H_3 hh-\fr 1 2 \la_6 \La H_4 hh$ respectively, with $\la' \equiv \la_4-2\sqrt{2}\fr{f}{w}\fr{uv}{u^2+v^2}$. The thermally-averaged annihilation cross-section times relative velocity is straightforwardly computed, yielding the relic density 
\be \Om h^2 \simeq 0.1\left(\fr{m_{H_7}}{1.354\ \mathrm{TeV}}\right)^2 \left(\bar{\la}-\fr{\la_5\la_6}{2(\la_{1\Xi}+\la_{2\Xi})}+\la'\fr{m_{H_3}^2}{4m^2_{H_7}-m^2_{H_3}}\right)^{-2}.\label{mhs}\ee We devide into two cases, 
\ben
\item $m_{H_7}\ll m_{H_3}$: Approximate 
\be \Om h^2 \simeq 0.1 \left(\fr{m_{H_7}}{\la_{\mathrm{eff}}\times 1.354\ \mathrm{TeV}}\right)^2,\ee where $\la_{\mathrm{eff}}\equiv \bar{\la}-\fr{\la_5\la_6}{2(\la_{1\Xi}+\la_{2\Xi})}-\la'$. Like $H_4$, the $H_3$ field is integrated out (i.e., both $H_{3,4}$ portals are not active), that all contribute to the contact interaction determined by the effective coupling $\la_{\mathrm{eff}}$. This case gives the correct abundance, if  
\be m_{H_7}\leq |\la_{\mathrm{eff}}|\times 1.354\ \mathrm{TeV}\sim 1.354\ \mathrm{TeV},\ \mathrm{for}\  |\la_{\mathrm{eff}}|\sim1.\ee Thus, the effective contact interaction predicts the dark matter mass bound in the range $m_{H_7} = 0.677$--$2.031\ \mathrm{TeV}$, for $|\la_{\mathrm{eff}}|=0.5$--$1.5$, respectively.   
\item $m_{H_7}\sim m_{H_3}$: After going beyond a viable low mass regime (somewhat similar to the previous case), the $H_7$ abundance is generally overpopulated, but having a resonance,
\be \Om h^2 \simeq 0.1 \left(\fr{m_{H_7}}{1.354\ \mathrm{TeV}}\right)^2\left(\fr{4m^2_{H_7}-m^2_{H_3}}{\la' m_{H_3}^2}\right)^2\rightarrow 0,\label{chdmd}\ee at $m_{H_7}=\fr 1 2  m_{H_3}=\fr 1 2 \sqrt{\bar{\la}'}w\simeq 2.6$ TeV, that again derives a correct relic density as desirable. Here, we have taken $w=9$~TeV that is fixed by the $\mathcal{Z}_1$ mass bound and $\bar{\la}' =1/3$ so that the relevant resonance exists below the regime $m_{Y_R}=g_Rw/2 \simeq 2.933$ TeV (assumed $g_R=g_L$). After the resonance, $m_{H_7}>\fr 1 2 m_{H_3}$, the density quickly rises as $\Om h^2 \sim m^2_{H_7}$, before it meets the WIMP unstable regime for $m_{H_7}>m_{Y_R}$. Let us remind the reader that the coannihilation processes such as $H_7 Y_R$ and $Y_R Y_R$---which happen when $m_{H_7}$ is close to $m_{Y_R}$---may significantly reduce the abundance, which is not considered. To be concrete, we plot the general density $\Om h^2$ given in (\ref{mhs}) as a function of the dark matter mass $m_{H_7}$ in Fig.~\ref{scalarH7} (curved line) for $m_{H_3}=5.2$ TeV, $\bar{\la}-\la_5\la_6/2(\la_{1\Xi}+\la_{2\Xi})=0.6$, and $\la'=1$. In order to fit the experimental density value $\Om h^2=0.11$ (shown in the figure as straight line) \cite{darkrv}, with the choice of parameter values, the $H_7$ mass varies beyond the weak scale up to 670~GeV and a region of resonance $1.75\ \mathrm{TeV}< m_{H_7}<2.933$ TeV, which encompasses the resonant point $\fr 1 2 m_{H_3}=2.6$ TeV and is bounded by $m_{Y_R}=2.933$ TeV.
\een 

Since $H_7$ is a standard model singlet, it only scatters off quarks via the Higgs portal $h$ (i.e., $t$-channel $h$-exchange), unlike the case of the scalar doublet $H_6$. The dark matter-nucleon scattering cross-section can easily be evaluated to be 
\be \sigma_{H_7-p,n}\simeq \left(\fr{2.6\ \mathrm{TeV}}{m_{H_7}}\right)^2\left(\fr{\bar{\la}}{0.65}\right)^2 3.88\times 10^{-45}\ \mathrm{cm}^2,\ee in agreement with \cite{ad5}. This prediction coincides with the direct detection limit from the XENON1T experiment $\sigma_{H_7-p,n}\sim 3.88\times 10^{-45}\ \mathrm{cm}^2$ at 90\% confidence level for the dark matter mass around the resonant point $m_{H_7}\sim 2.6$ TeV and the sizable Higgs-portal coupling $\bar{\la}\sim 0.65$~\cite{Xenon}. With such $\bar{\la}$ fixed, the lower mass regions of ${H_7}$ (including case 1 and low mass regime of case 2) should be ruled out by the direct detection. We would like to stress that the dark matter mass in considered model is in few TeVs even larger than that, where the abundance is governed by the new physics behind. An indirect detection is very insignificant \cite{indirect} and is ignored in this work. 

The last remark is that from (\ref{chdmd}) we obtain the resonance width (neglect the bound $m_{Y_R}$) proportional to $\la'\times 1.354$ TeV, which is large due to the large $\la'=1$ (as taken), in contrast to the case of fermion dark matter governed by the heavy gauge portal. 
	
\subsubsection{Remark on gauge boson dark matter}

We would like to emphasize that the mass of the vector gauge boson $X_R^0$ is radically larger than that of the 
vector gauge boson $Y_R^\pm$, as we see from \cite{2L3R} and above that $m_{X_R}\simeq g_R\La/\sqrt{2}\gg m_{Y_R}\simeq g_R w/2$ for every $g_R$. So, the vector gauge boson $X^0$ cannot be a dark matter candidate since it is unstable, entirely decaying into the $Y_R^\pm$ and standard model gauge bosons ($W^\mp$).
  
\subsection{The 3-2-3-1 model with $q=-1$}

In this model, the colorless and neutral wrong particles are the scalar $H_8^0$ and gauge boson $Y_R^0$.  
First, we assume that the vector field $Y_R^0$ is a LWP.  It directly couples to the $W^\pm$, $Z$ gauge bosons, and the dominated annihilation channels are $Y^0_R Y^{0*}_R \rightarrow W^+W^-$, $ZZ$.  The dark
matter thermal relic abundance is approximated as 
\bea
\Omega_{Y_R} h^2 \simeq 10^{-3}\fr{m^2_{W}}{m^2_{Y_R}}.
\eea 
Because the fraction $\fr{m_W^2}{m_{Y_R}^2}$ is very small, their relic abundance is $\Omega_{Y_R} h^2 \ll 10^{-3}$, much lower than that measured by WMAP/PLANCK \cite{darkrv}. 

This under-abundance may be evaded by signifying that the vector candidate is superheavy, and non-thermally created as associated with the reheating process or by the gravitational mechanism. Here, the dark matter is never to thermalize but it derives a corrected relic abundance \cite{Superdarkmatter}. 

Next, the scalar field, $H_8^0$, is considered as a LWP. Because it transforms as the doublet of $SU(2)_L$ group, it directly couples to the standard model gauge boson and behaves like the $H_6^0$ scalar field, see in Sec. \ref{SDM}. Hence, we have not repeated it here. 

To conclude this section, we have focussed primarily on the dark matter abundances and direct detections. Since our candidates are heavy, the indirect detections as well as the current collider searches are insignificant. But, when the LHC is run at $\sqrt{s}=14$ TeV with high integrated luminosity, it is worth searching for.  

\section{\label{Con}Conclusions}

Unlike the minimal left-right symmetric model, the $SU(3)_C \otimes SU(2)_L \otimes SU(3)_R \otimes U(1)_X$ model treats the baryon-minus-lepton number as a non-Abelian gauge charge, analogous to the electric charge, which provides a nontrivial
unification framework for the electroweak and $B-L$ interactions as well as manifestly unifying the dark (wrong $B-L$) and normal sectors in gauge multiplets. The matter parity $W_P=(-1)^{3(B-L)+2s}$ is a residual gauge symmetry, transforming nontrivially on the dark fields. The conservation of $W_P$ means that the lightest wrong $B-L$ particle is stabilized, responsible for dark matter. The electric charge parameter (i.e., the electric charge of $E_a$) is constrained by $-1.822<q<0.822$. If the new leptons $E_a$ carry integer charges, there exist two dark matter models corresponding to $q=0$ and $q=-1$. These dark matter models always have the Landau poles larger than the Planck scale. 

The neutrino masses are naturally induced by a seesaw mechanism. Since the Dirac neutrino masses are related to those of the charged leptons, the seesaw scale ranges from $10^4$ GeV or $10^{16}$ GeV depending on the weak scale ratio $u/v$. At the low seesaw scale, the lepton flavor violation decays $\mu \rightarrow 3e$ and $\mu \rightarrow e \gamma$ are dominantly induced by a doubly-charged Higgs exchange. The decay rates are consistent with the experimental bounds if the doubly-charged Higgs mass varies from few TeVs to hundred TeVs. 

The model contains two new neutral gauge bosons $\mathcal{Z}_1, \mathcal{Z}_1^\prime$ in which $\mathcal{Z}_1^\prime$ has mass at the seesaw scale, more heavier than $\mathcal{Z}_1$. Thus, the field $\mathcal{Z}_1$ is accessible at the colliders as well as governing the dark matter observable, unlike $\mathcal{Z}'_1$. The LEPII constrains the $\mathcal{Z}_1$ mass at $\mathcal{O}(1)$ TeV, while the LHC searches show that the $\mathcal{Z}_1$ mass is larger than 4 TeV for $\sqrt{s}=13$ TeV.

We investigate the two viable dark matter models. The model $q=0$ contains two types of dark matter, fermion and scalar fields. The fermion dark matter relics  is dominated by the $\mathcal{Z}_1$ gauge boson in every symmetry breaking scheme. There always exits a resonance $\mathcal{Z}_1$ and narrow region for the dark matter mass that produces the correct abundance, in agreement with the $\mathcal{Z}_1$ bounds. The scalar dark matter can be a $SU(2)_L$ doublet or a $SU(2)_L$ singlet, which both can reproduce the correct relic density. But the doublet candidate may be ruled out by the direct detection experiments. The model $q=-1$ also contains two kinds of dark matter. The scalar $SU(2)_L$ doublet candidate behaves similarly to the scalar doublet in the previous model, and thus ruled out. The vector candidate is stabilized, but has a thermal abundance far bellow the WMAP/PLANCK predictions. In short, the two models predict distinct scenarios for dark matter.  

\section*{Acknowledgement}

NTN thanks Farinaldo S. Queiroz for the helpful discussions. This research is funded by Vietnam National Foundation for Science and Technology Development (NAFOSTED) under grant number 103.01-2017.05. We acknowledge the financial support of the International Centre of Physics at the Institute of Physics, Vietnam Academy of Science and Technology.


\begin{thebibliography}{99}

\bibitem{neuos} T. Kajita, {\it Nobel Lecture: Discovery of atmospheric neutrino oscillations}, Rev. Mod. Phys. {\bf 88}, 030501 (2016); A. B. McDonald, {\it Nobel Lecture: The Sudbury Neutrino Observatory: Observation of flavor change for solar neutrinos},
Rev. Mod. Phys. {\bf 88}, 030502 (2016).

\bibitem{darkrv} D. N. Spergel {\it et al.} (WMAP Collaboration), Astrophys. J. Suppl. Ser. {\bf 170}, 377 (2007); P. A. R. Ade {\it et al.} (Planck Collaboration), Astron. Astrophys. {\bf 571}, A1 (2014); See, for reviews, G. Bertone, D. Hooper, and J. Silk, Phys. Rep. {\bf 405}, 279 (2005); G. Jungman, M. Kamionkowski, and K. Griest, Phys. Rep. {\bf 267}, 195 (1996).

\bibitem{data} C. Patrignani {\it et al.} (Particle Data Group), Chin. Phys. C {\bf 40}, 100001 (2016), and partial updates at http://pdg.lbl.gov.

\bibitem{LR1} J. C. Pati and A. Salam, Phys. Rev. D \textbf{10}, 275 (1974); R. N. Mohapatra and J. C. Pati,
Phys. Rev. D \textbf{11}, 566 (1975); R. N. Mohapatra and J. C. Pati, Phys. Rev. D \textbf{11}, 2558
(1975); G. Senjanovi\'c and R. N. Mohapatra, Phys. Rev. D \textbf{12}, 1502 (1975); G. Senjanovi\'c, Nucl.
Phys. B \textbf{153}, 334 (1979).

\bibitem{NLR} P. Minkowski, Phys. Lett. B \textbf{67}, 421 (1977);
 R. N. Mohapatra and G. Senjanovi\'c, Phys. Rev. Lett. \textbf{44}, 912 (1980);  R. N. Mohapatra and G. Senjanovi\'c,
Phys. Rev. D \textbf{23}, 165 (1981).

\bibitem{ad0} J. Schechter and J. W. F. Valle, Phys. Rev. D {\bf 22}, 2227 (1980); Phys. Rev. D {\bf 25}, 774 (1982).  

\bibitem{PheLR} G. Beall, M. Bander, and A. Soni, Phys. Rev. Lett. \textbf{48}, 848
(1982), R. N. Mohapatra, G. Senjanovich, and M. Tran, Phys. Rev. D \textbf{28}, 546 (1983); G. Ecker,
W. Grimus, and H. Neufeld, Phys. Lett. B \textbf{127}, 365 (1983);  F. G. Gilman and M. H.
Reno, Phys. Lett. B \textbf{127}, 426 (1983); F. G. Gilman and M. H. Reno, Phys. Rev. D \textbf{29},
937 (1983), G. Ecker and W. Grimus, Nucl. Phys. B \textbf{258}, 328 (1985); J. -M. Frere {\it et al.}, Phys. Rev. D {\bf 46}, 337 (1992); 
M. E. Pospelov, Phys. Rev. D \textbf{56}, 259 (1997) [arXiv:hep-ph/9611422]; A. Maiezza {\it et al.}, Phys. Rev. D {\bf 82}, 055022 (2010) [arXiv:1005.5160 [hep-ph]]. 

\bibitem{vubpro} A. Crivellin, Phys. Rev. D {\bf 81}, 031301 (2010) [arXiv:0907.2461 [hep-ph]]; A. J. Buras, K. Gemmler, and G. Isidori, Nucl. Phys. B {\bf 843}, 107 (2011) [arXiv:1007.1993 [hep-ph]]; M. Blanke, A. J. Buras, K. Gemmler, and T. Heidsieck, JHEP {\bf 03}, 024 (2012) [arXiv:1111.5014 [hep-ph]].

\bibitem{ad1} F. Bezrukov, H. Hettmansperger, and M. Lindner, Phys. Rev. D {\bf 81}, 085032 (2010); M. Nemevsek, G. Senjanovic, and Y. Zhang, JCAP {\bf 07}, 006 (2012); J. Barry, J. Heeck, and W. Rodejohann, JHEP {\bf 07}, 081 (2014). 

\bibitem{ad2} J. Heeck and S. Patra, Phys. Rev. Lett. {\bf 115}, 121804 (2015); C. Garcia-Cely and J. Heeck, JCAP {\bf 03}, 021 (2016); M. Cirelli, N. Fornengo, and A. Strumia, Nucl. Phys. B {\bf 753}, 178 (2006); A. Berlin, P.J. Fox, D. Hooper, and G. Mohlabeng, JCAP {\bf 06}, 016 (2016); P.S.B. Dev, R.N. Mohapatra, and Y. Zhang, JHEP {\bf 11}, 077 (2016).

\bibitem{3311dm} P. V. Dong, T. D. Tham, and H. T. Hung, Phys. Rev. D {\bf 87}, 115003 (2013) [arXiv:1305.0369 [hep-ph]]; P. V. Dong, D. T. Huong, F. S. Queiroz, and N. T. Thuy, Phys. Rev. D {\bf 90}, 075021 (2014) [arXiv:1405.2591 [hep-ph]]; D. T. Huong, P. V. Dong, C. S. Kim, and N. T. Thuy, Phys. Rev. D {\bf 91}, 055023 (2015) [arXiv:1501.00543 [hep-ph]]; P. V. Dong, Phys. Rev. D {\bf 92}, 055026 (2015) [arXiv:1505.06469 [hep-ph]]; P. V. Dong and D. T. Si, Phys. Rev. D {\bf 93}, 115003 (2016) [arXiv:1510.06815 [hep-ph]]; A. Alves, G. Arcadi, P. V. Dong, L. Duarte, F. S. Queiroz, and J. W. F. Valle, Phys. Lett. B 772, 825 (2017) [arXiv:1612.04383 [hep-ph]].   

\bibitem{LR331} A. G. Dias, C. A. de S. Pires, and P. S. Rodrigues da Silva, Phys. Rev. D {\bf 82}, 035013 (2010); C. P. Ferreira, M. M. Guzzo, and P. C. de Holanda, Braz. J. Phys. {\bf 46}, 453 (2016) [arXiv:1509.02977 [hep-ph]].

\bibitem{3L3R} D. T. Huong and P. V. Dong, Phys. Rev. D {\bf 93}, 095019 (2016) [arXiv:1603.05146 [hep-ph]].

\bibitem{2L3R}P. V. Dong, D. T. Huong, D. V. Loi, N. T. Nhuan, N. T. K. Ngan, Phys. Rev. D {\bf 95}, 075034 (2017) [arXiv:1609.03444 [hep-ph]].

\bibitem{3L3RVE} P. V. Dong and D. T. Huong, Commun. Phys. {\bf 28}, 21 (2018) [arXiv:1610.02642 [hep-ph]]; M. Reig, J. W. F. Valle, C. A. Vaquera-Araujo, Phys. Lett. B {\bf 766}, 35 (2017) [arXiv:1611.02066 [hep-ph]]; M. Reig, J. W. F. Valle, C. A. Vaquera-Araujo, JHEP {\bf 05}, 100 (2017) [arXiv:1611.04571 [hep-ph]]; C. Hati, S. Patra, M. Reig, J. W. F. Valle, and C. A. Vaquera-Araujo, Phys. Rev. D {\bf 96}, 015004 (2017) [arXiv:1703.09647 [hep-ph]]; P. V. Dong, D. T. Huong, Farinaldo S. Queiroz, J. W. F. Valle, and C. A. Vaquera-Araujo, JHEP {\bf 04}, 143 (2018) [arXiv:1710.06951 [hep-ph]]. 

\bibitem{lrema} C. Kownacki, E. Ma, N. Pollard, O. Popov, and M. Zakeri, Phys. Lett. B {\bf 777}, 121 (2018) [arXiv:1710.00762 [hep-ph]]; E. Ma, arXiv:1712.08994 [hep-ph]; C. Kownacki, E. Ma, N. Pollard, O. Popov, and M. Zakeri, Nucl. Phys. B {\bf 928}, 520 (2018).

\bibitem{331} F. Pisano and V. Pleitez, Phys. Rev. D {\bf 46}, 410 (1992) [arXiv:hep-ph/9206242]; P. H. Frampton, Phys. Rev. Lett. {\bf 69}, 2889
(1992); R. Foot, O. F. Hernandez, F. Pisano, and V. Pleitez, Phys. Rev. D {\bf 47}, 4158 (1993) [arXiv:hep-ph/9207264],
M. Singer, J. W. F. Valle, and J. Schechter, Phys. Rev. D {\bf 22}, 738 (1980); J. C. Montero, F. Pisano,
and V. Pleitez, Phys. Rev. D {\bf 47}, 2918 (1993); R. Foot, H. N. Long, and Tuan A. Tran, Phys. Rev. D
{\bf 50}, 34 (1994) [arXiv:hep-ph/9402243]; P. V. Dong, H. N. Long, D. T. Nhung, and D. V. Soa, Phys. Rev. D {\bf 73}, 035004 (2006) [arXiv:hep-ph/0601046];
S. M. Boucenna, J. W. F. Valle, and A. Vicente, Phys. Rev. D {\bf 92}, 053001 (2015) [arXiv:1502.07546 [hep-ph]]; J. W. F. Valle and
C. A. Vaquera-Araujo, Phys. Lett. B {\bf 755}, 363 (2016) [arXiv:1601.05237 [hep-ph]].

\bibitem{bsmumu} V. Khachatryan {\it et al.} (CMS and LHCb Collaborations), Nature {\bf 522}, 68 (2015) [arXiv:1411.4413 [hep-ex]]; R. Aaij {\it et al.} (LHCb Collaboration), JHEP {\bf 02}, 104 (2016); R. Aaij {\it et al.} (LHCb Collaboration), JHEP {\bf 09}, 179 (2015) [arXiv:1506.08777 [hep-ex]]; R. Aaij {\it et al.} (LHCb Collaboration), Phys. Rev. Lett. {\bf 113}, 151601 (2014) [arXiv:1406.6482 [hep-ex]]; S. Descotes-Genon {\it et al.}, JHEP {\bf 06}, 092 (2016) [arXiv:1510.04239 [hep-ph]].     

\bibitem{muegamma1} T.-P. Cheng and L.-F. Li, Phys. Rev. D {\bfseries 44}, 1502 (1991); B. He, T.-P. Cheng, L.-F. Li, Phys. Lett. B {\bf 553}, 277 (2003) [arXiv:hep-ph/0209175].

\bibitem{muegamma2} M. Kakizaki, Y. Ogura, and F. Shima, Phys. Lett. B {\bf 566}, 210 (2003) [arXiv:hep-ph/0304254]; P. V. Dong, H. N. Long, Phys. Rev. D {\bf 77}, 057302 (2008) [arXiv:0801.4196 [hep-ph]].

\bibitem{LEP} The LEP collaborations: ALEPH collaboration, DELPHI collaboration, L3 collaboration, OPAL collaboration, the LEP electroweak working group, arXiv:hep-ex/0612034; M. Carena, A. Daleo, B. A. Dobrescu, and T. M. P. Tait, Phys. Rev. D {\bf70}, 093009 (2004) [arXiv:hep-ph/0408098]. 

\bibitem{ALA} ATLAS Collaboration, ATLAS-CONF-2016-045, 2016.

\bibitem{ALA1} E. Accomando, A. Belyaev, L. Fedeli, S. F. King, C. Shepherd-Themistocleous, Phys. Rev. D {\bf 83}, 075012 (2011) [arXiv:1010.6058 [hep-ph]]. 

\bibitem{ALA2} A. D. Martin, W. J. Stirling, R. S. Thorne, and G. Watt, Eur. Phys. J. C {\bf 63}, 189 (2009) [arXiv:0901.0002 [hep-ph]]; See also:
 V. Bertone, S. Carrazza, D. Pagani, and M. Zaro, JHEP {\bf 11}, 194 (2015) [arXiv:1508.07002 [hep-ph]]; C. Buttar {\it et al.}, {\it Les Houches Physics at TeV Colliders 2005, Standard Model and Higgs working group: Summary report}, arXiv:hep-ph/0604120. 

\bibitem{Alast2017} M. Aaboud {\it et al.} (ATLAS Collaboration), JHEP {\bf10}, 182 (2017) [arXiv:1707.02424 [hep-ex]].

\bibitem{Xenon} E. Aprile { \it et al.}, (XENON Collaboration), Phys. Rev. Lett. {\bf 119}, 181301 (2017). 

\bibitem{inertdoublet} L. L. Honorez, E. Nezri, J. F. Oliver, and M. H. G. Tytgat, JCAP {\bf 0702}, 028 (2007) [arXiv:hep-ph/0612275]; M. Gustafsson, E. Lundstrom, L. Bergstrom, and J. Edsjo, Phys. Rev. Lett. {\bf 99}, 041301 (2007) [arXiv:astro-ph/0703512]; E. Ma, Phys.
Rev. D {\bf 73}, 077301 (2006) [arXiv:hep-ph/0601225]; R. Barbieri, L. J. Hall, and V. S. Rychkov, Phys. Rev. D {\bf 74}, 015007 (2006) [arXiv:hep-ph/0603188]. 

\bibitem{indirect} B. S. Acharya {\it et al.} (Cherenkov Telescope Array Consortium Collaboration), arXiv:1709.07997 [astro-ph.IM].

\bibitem{ad5} P. V. Dong, C. S. Kim, N. T. Thuy, and D. V. Soa, Phys. Rev. D {\bf 91}, 115019 (2015).  

\bibitem{Superdarkmatter} D. T. Huong and P. V. Dong, Eur. Phys. J. C {\bf 77}, 204 (2017) [arXiv:1605.01216 [hep-ph]]. 

\end{thebibliography}
 \end{document}